\documentclass[12pt]{article}
\usepackage{epsfig}
\usepackage{graphicx}
\usepackage{amsmath}

\begin{document}

 \newcommand{\beq}{\begin{equation}}
\newcommand{\eeq}{\end{equation}}
\newcommand{\bea}{\begin{eqnarray}}
\newcommand{\eea}{\end{eqnarray}}
\newcommand{\beqn}{\begin{eqnarray}}
\newcommand{\eeqn}{\end{eqnarray}}
\newcommand{\beas}{\begin{eqnarray*}}
\newcommand{\eeas}{\end{eqnarray*}}
\newcommand{\defi}{\stackrel{\rm def}{=}}
\newcommand{\non}{\nonumber}
\newcommand{\bquo}{\begin{quote}}
\newcommand{\enqu}{\end{quote}}
\newcommand{\qt}{\tilde q}

\newcommand{\gsim}{\lower.7ex\hbox{$
\;\stackrel{\textstyle>}{\sim}\;$}}
\newcommand{\lsim}{\lower.7ex\hbox{$
\;\stackrel{\textstyle<}{\sim}\;$}}


\def\de{\partial}
\def\Tr{ \hbox{\rm Tr}}
\def\const{\hbox {\rm const.}}
\def\o{\over}
\def\im{\hbox{\rm Im}}
\def\re{\hbox{\rm Re}}
\def\bra{\langle}\def\ket{\rangle}
\def\Arg{\hbox {\rm Arg}}
\def\Re{\hbox {\rm Re}}
\def\Im{\hbox {\rm Im}}
\def\diag{\hbox{\rm diag}}


\def\QATOPD#1#2#3#4{{#3 \atopwithdelims#1#2 #4}}
\def\stackunder#1#2{\mathrel{\mathop{#2}\limits_{#1}}}
\def\stackreb#1#2{\mathrel{\mathop{#2}\limits_{#1}}}
\def\Tr{{\rm Tr}}
\def\res{{\rm res}}
\def\Bf#1{\mbox{\boldmath $#1$}}
\def\balpha{{\Bf\alpha}}
\def\bbeta{{\Bf\beta}}
\def\bgamma{{\Bf\gamma}}
\def\bnu{{\Bf\nu}}
\def\bmu{{\Bf\mu}}
\def\bphi{{\Bf\phi}}
\def\bPhi{{\Bf\Phi}}
\def\bomega{{\Bf\omega}}
\def\blambda{{\Bf\lambda}}
\def\brho{{\Bf\rho}}
\def\bsigma{{\bfit\sigma}}
\def\bxi{{\Bf\xi}}
\def\bbeta{{\Bf\eta}}
\def\d{\partial}
\def\der#1#2{\frac{\d{#1}}{\d{#2}}}
\def\Im{{\rm Im}}
\def\Re{{\rm Re}}
\def\rank{{\rm rank}}
\def\diag{{\rm diag}}
\def\2{{1\over 2}}
\def\ntwo{${\mathcal N}=2\;$}
\def\nfour{${\mathcal N}=4\;$}
\def\none{${\mathcal N}=1\;$}
\def\x{\stackrel{\otimes}{,}}

\def\ba{\beq\new\begin{array}{c}}
\def\ea{\end{array}\eeq}
\def\be{\ba}
\def\ee{\ea}
\def\stackreb#1#2{\mathrel{\mathop{#2}\limits_{#1}}}

\def\Tr{{\rm Tr}}
\newcommand{\vp}{\varphi}
\newcommand{\pt}{\partial}
\newcommand{\ve}{\varepsilon}
\renewcommand{\theequation}{\thesection.\arabic{equation}}

\setcounter{footnote}0

\vfill

\begin{titlepage}

\begin{flushright}
FTPI-MINN-09/02, UMN-TH-2733/09\\
January 26, 2009
\end{flushright}

\begin{center}
{  \Large \bf  Crossover between  Abelian\\[0.5mm]
 and non-Abelian confinement
in
\\[3mm]
 \boldmath{\ntwo} supersymmetric QCD}
\end{center}

\vspace{1mm}

\begin{center}

 {\large
 \bf    M.~Shifman$^{\,a}$ and \bf A.~Yung$^{\,\,a,b,c}$}
\end {center}

\begin{center}


$^a${\it  William I. Fine Theoretical Physics Institute,
University of Minnesota,
Minneapolis, MN 55455, USA}\\
$^{b}${\it Petersburg Nuclear Physics Institute, Gatchina, St. Petersburg
188300, Russia\\
$^c${\it Institute of Theoretical and Experimental Physics, Moscow
117259, Russia}}
\end{center}

\begin{center}
{\large\bf Abstract}
\end{center}
In this paper we investigate the nature
 of the transition from Abelian to non-Abelian confinement
 (i.e. crossover vs. phase transition).
To this end we
consider the basic \ntwo model where non-Abelian flux tubes (strings)
were first found:  supersymmetric QCD with the U($N$) gauge group
and $N_f=N$ flavors of fundamental matter (quarks). The Fayet--Iliopoulos
term $\xi$ triggers the squark condensation and leads to  
the formation of non-Abelian strings.
There are two adjustable parameters in this model: $\xi$ and the quark mass
difference $\Delta m$. We obtain the phase diagram on the
 ($\xi,\, \Delta m$) plane.
At large $\xi$ and small $\Delta m$  the worldsheet
 dynamics
of  the string  orientational moduli   is described by 
\ntwo two-dimensional CP$(N-1)$ model.
 We show that as we reduce $\xi$ the theory exhibits a
crossover to the Abelian (Seiberg--Witten) regime. Instead of $N^2$ 
degrees of freedom of
non-Abelian theory now only $N$ degrees of freedom survive
 in the low-energy spectrum.
Dyons with certain quantum numbers condense leading to the formation of
the Abelian $Z_N$ strings whose fluxes are fixed inside the Cartan 
subalgebra of the gauge group. As we increase $N$ this crossover 
becomes exceedingly sharper becoming   a genuine phase transition
at  $N =\infty$.

\vspace{2cm}

\end{titlepage}

\newpage


\section {Introduction}
\label{intro}
\setcounter{equation}{0}

Transition from Abelian to non-Abelian confinement emerged as 
a central question in the current explorations of Yang--Mills theories.
By Abelian confinement we mean that only those gauge bosons that lie in the Cartan subalgebra
are dynamically important in the infrared, i.e. at distances of the order of the inverse 
flux tube (string) size.
By  non-Abelian
confinement we mean such dynamical regime in which at distances
of the flux tube formation all gauge bosons are equally important.
In supersymmetric \ntwo Yang--Mills theories slightly deformed by a
$\mu\Tr\Phi^2$ term linear confinement was discovered \cite{SW1}, explained 
by the dual Meissner effect.  
In the limit of small $\mu$ amenable to analytic studies
\cite{SW1} confinement is  Abelian. It is believed that as $\mu$ gets
large, $\mu\gsim \Lambda$, a smooth transition to non-Abelian confinement takes place
in the Seiberg--Witten model.

In non-supersymmetric theories a similar purpose construction, with an adjustable parameter,
was engineered in \cite{SU2} (see also \cite{SU3}) where Yang--Mills theories on 
$R_3\times S_1$ were considered. The radius of the compact dimension $r(S_1)$
was treated as a free parameter. 
At small  $r(S_1)$, after a center-symmetric stabilization, linear Abelian confinement
sets in by virtue of the Polyakov mechanism \cite{Pol}.  
Then it was argued that the transition 
from the small-$r(S_1)$ Abelian confinement regime to the
decompactification limit of large $r$, $r(S_1)\gg \Lambda^{-1}$,
where confinement is non-Abelian, is smooth. No obvious order parameter
that could discontinuously change in passing from small to large $r(S_1)$ was detected.

This paper presents our new results on this issue. The nature of the transition from the 
Abelian to non-Abelian regime -- i.e. phase transition vs. crossover -- appears to be non
 universal.
If there is a discrete symmetry on the string world sheet and the mode of realization of this
 symmetry changes in passing
from the Abelian to non-Abelian regime then these two domains are separated by a phase
 transition.
A particular nonsupersymmetric example \cite{GSY05,GSYphtr} of such a situation will be discussed
in the bulk of the paper.

On the other hand, if the mode of realization of the discrete symmetry does not change, 
or there is no appropriate symmetry whatsoever, then the Abelian confinement is separated from the
non-Abelian one by a crossover rather than phase transition.
 In this paper we focus on
\ntwo Yang--Mills theory with the gauge group U($N$) and $N_f=N$ flavors.
To simplify our discussion we  mostly consider the $N=2$ case of U(2) theory with two flavors.
We will show that this model belongs to the second class, with smooth transitions of the crossover type. Although
there is no phase transition,  we find that both 
perturbative and non-perturbative
low energy spectra of the theory are drastically changed when
we pass from the Abelian to non-Abelian regime.

The benchmark model we will deal with --  the U(2) theory with $N_f=2$ (s)quark multiplets
-- is described in detail in the review paper \cite{SYrev}. 
It is worth recalling that the model is characterized by two adjustable parameters:
the coefficient of the Fayet--Iliopoulos (FI) term $\xi$ and the difference $\Delta m$
of the mass terms of the first and second flavors. If $\xi\gg \Lambda^2 $,
where $\Lambda$ is the dynamical scale of the gauge theory at hand, the theory is at weak 
coupling
and can be exhaustively analyzed using quasiclassical methods. The domain of large $|\Delta m |$
is that of the Abelian confinement. At small $|\Delta m |$ confinement is non-Abelian.
A discrete $Z_2$ symmetry inherent to the Lagrangian of the world-sheet theory
is spontaneously broken in both limits, 
albeit the order parameters are different. Thus we expect (and, in  fact, demonstrate)
a crossover in  $|\Delta m |$. 

\begin{figure}
\epsfxsize=7cm
\centerline{\epsfbox{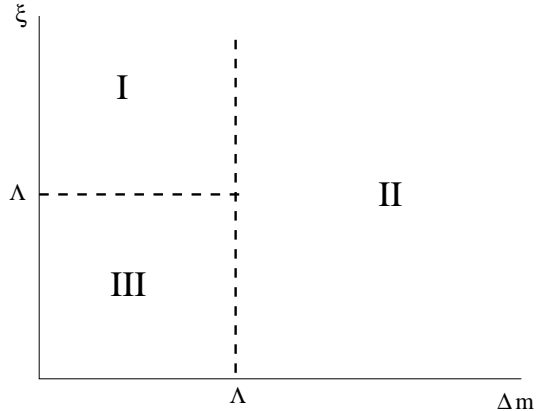}}
\caption{\small Various regimes in the benchmark \ntwo
model  are  separated by crossovers.}
\label{figphasediag}
\end{figure}

The domain of small $\xi$ was not considered previously in the 
context of the problem we pose. Our task is to
include it in consideration.
Various regimes  of the theory in the $(\xi,\,\Delta m)$ plane are schematically shown in Fig.~\ref{figphasediag}. We choose $\Delta m$ real, which is always possible to achieve by an appropriate U(1) rotation.
The vertical axis in this figure denotes the values of the FI parameter $\xi$ while the horizontal
axis  represents the quark mass difference.

As was mentioned, the domain I is that of non-Abelian confinement. 
In this domain the perturbative spectrum of the bulk 
theory has $N^2$ light states.\footnote{By light we mean those states
whose masses are less than or of the order of the inverse size of the string.} In  the limit of 
degenerate quark masses
the bulk theory has an unbroken global SU(2)$_{C+F}$ symmetry (the so-called
color-flavor locking,
see Sect.~\ref{bulk}), and the light states come in adjoint and singlet 
representations of this group. The nonperturbative spectrum contains
mesons built from monopole-antimonopole pairs connected by two strings,
\cite{SYrev}. These strings are 
non-Abelian \cite{HT1,ABEKY,SYmon,HT2}
(see also the reviews \cite{Trev,SYrev,Jrev,Trev2}). 
The non-Abelian SU(2) part of their fluxes is determined
by moduli parameters, whose dynamics is described by \ntwo 
supersymmetric CP(1)  on the string world sheet. Due to large quantum fluctuations
in the CP(1)  model the average non-Abelian flux of such a string vanishes.

The domain  II is that of  Abelian confinement at weak coupling. As we increase  
  $\Delta m$, the $W$ bosons and their superpartners become
heavy and decouple from the low-energy spectrum. We are  left with two
photon states and their quark \ntwo superpartners. Strings also become
Abelian $Z_2$ strings. The moduli of  the CP(1)  model at large  $\Delta m$
are fixed in two definite directions -- the north and south poles
of the $S_2$ sphere (the target space of the CP(1) sigma model is $S_2$). 
The non-Abelian parts of their fluxes no longer vanish.

As we reduce $\xi$ and  $|\Delta m|$ we enter the domain III. It is nothing but the Abelian
Seiberg--Witten confinement \cite{SW1,SW2}. The $W$ bosons and their superpartners
decay on the curves of the marginal stability (CMS). The set of the light states includes
photons and dyons with certain quantum numbers (the quarks we started with become dyons
due to monodromies as we reduce $|\Delta m|$). 
Condensation of dyons leads
to formation of Abelian $Z_2$ strings.  Nonperturbative
spectrum still contains mesons built of monopole-antimonopole pairs
connected by two distinct $Z_2$ strings. However, now these strings
are different from those in the domain I. They have nonvanishing fluxes
directed in  the Cartan subalgebra of SU(2).

These three regimes are separated by crossover transitions. 
Thus, non-Abelian strings can smoothly evolve into Abelian ones and vice versa.
This is our main finding.

If, instead of the benchmark U(2) model we considered its U$(N)$ generalization, 
we would see that, 
as we increase the number of colors
 $N$, these crossovers 
become exceedingly  sharper and transform into genuine phase transitions
in the limit $N\to\infty$. 
The width of the crossover domain scales as $1/N$.
At finite $N$ there is no discontinuity
in physical observables (and associated breaking of relevant global symmetries)
along the dashed lines in Fig.~\ref{figphasediag}. Still both the perturbative
spectrum and confining strings are dramatically different in the regimes I, II and III.
 In particular, if we keep the quark mass difference  $\Delta m =0$
and reduce the FI parameter $\xi$ we pass from the regime with non-Abelian 
strings and non-Abelian monopole confinement to  the regime with 
 Abelian strings and Abelian confinement. The low-energy spectrum of the theory in, say, domain I is not mapped onto that in domain III.

The paper is organized as follows. In Sect.~\ref{bulk} we describe our bulk theory
at large $\xi$. In Sect.~\ref{3three} we make some preliminary remarks on the behavior of the theory
at small $\xi$. In Sect.~\ref{strings} we briefly review non-Abelian strings
and discuss the order parameter which can separate a non-Abelian string from Abelian one.
In Sect. \ref{coulombbranch} we describe our theory in domain III, 
while in Sect.~\ref{largeN} we consider the 
limit $N\to\infty$. Section~\ref{7seven} summarizes our findings.
In Appendix we present in more detail the CP$(N-1)$ model with twisted
masses and $Z_{2N}$ global symmetry.

\section{The bulk theory:  large values of the \\ FI parameter}
\label{bulk}
\setcounter{equation}{0}

In this section for convenience of the reader we will briefly outline
some basic features of the \ntwo bulk theory we will work with.
Since these features are general, we will assume the gauge group to be U$(N)$,
and only later we will set $N=2$.

We introduce $N$ flavors (each of which is described by two \none super\-fields, 
$Q$ and $\tilde Q$).
The field content 
 is as follows. The \ntwo vector multiplet
consists of the  U(1)
gauge field $A_{\mu}$ and the SU$(N)$  gauge field $A^a_{\mu}$,
where $a=1,..., N^2-1$, and their Weyl fermion superpartners
 plus
complex scalar fields $a$, and $a^a$. The latter are in the adjoint
representation of SU$(N)$.

The quark multiplets of  the SU$(N)\times$U(1) theory consist
of   the complex scalar fields
$q^{kA}$ and $\tilde{q}_{Ak}$ (squarks) and
their   fermion superpartners, all in the fundamental representation of 
the SU$(N)$ gauge group.
Here $k=1,..., N$ is the color index
while $A$ is the flavor index, $A=1,..., N$.
Note that the scalars $q^{kA}$ and ${\bar{\tilde q}}^{\, kA}$
form a doublet under the action of the   global
SU(2)$_R$ group.

The bosonic part of the bulk 
theory has the form \cite{ABEKY} (see also the review paper \cite{SYrev})
\beqn
S&=&\int d^4x \left[\frac1{4g^2_2}
\left(F^{a}_{\mu\nu}\right)^2 +
\frac1{4g^2_1}\left(F_{\mu\nu}\right)^2
+
\frac1{g^2_2}\left|D_{\mu}a^a\right|^2 +\frac1{g^2_1}
\left|\partial_{\mu}a\right|^2 \right.
\nonumber\\[4mm]
&+&\left. \left|\nabla_{\mu}
q^{A}\right|^2 + \left|\nabla_{\mu} \bar{\tilde{q}}^{A}\right|^2
+V(q^A,\tilde{q}_A,a^a,a)\right]\,.
\label{model}
\eeqn
Here $D_{\mu}$ is the covariant derivative in the adjoint representation
of  SU$(N)$, while
\beq
\nabla_\mu=\partial_\mu -\frac{i}{2}\; A_{\mu}
-i A^{a}_{\mu}\, T^a\,.
\label{defnabla}
\eeq
We suppress the color  SU($N$)  indices. The normalization of the 
 SU($N$) generators  $T^a$ is as follows
$$
{\rm Tr}\, (T^a T^b)=\mbox{$\frac{1}{2}$}\, \delta^{ab}\,.
$$
The coupling constants $g_1$ and $g_2$
correspond to the U(1)  and  SU$(N)$  sectors, respectively.
With our conventions, the U(1) charges of the fundamental matter fields
are $\pm1/2$.

The potential $V(q^A,\tilde{q}_A,a^a,a)$ in the action (\ref{model})
is the  sum of  $D$ and  $F$  terms,
\beqn
V(q^A,\tilde{q}_A,a^a,a) &=&
 \frac{g^2_2}{2}
\left( \frac{1}{g^2_2}\,  f^{abc} \bar a^b a^c
 +
 \bar{q}_A\,T^a q^A -
\tilde{q}_A T^a\,\bar{\tilde{q}}^A\right)^2
\nonumber\\[3mm]
&+& \frac{g^2_1}{8}
\left(\bar{q}_A q^A - \tilde{q}_A \bar{\tilde{q}}^A -N \xi\right)^2
\nonumber\\[3mm]
&+& 2g^2\left| \tilde{q}_A T^a q^A \right|^2+
\frac{g^2_1}{2}\left| \tilde{q}_A q^A  \right|^2
\nonumber\\[3mm]
&+&\frac12\sum_{A=1}^N \left\{ \left|(a+\sqrt{2}m_A +2T^a a^a)q^A
\right|^2\right.
\nonumber\\[3mm]
&+&\left.
\left|(a+\sqrt{2}m_A +2T^a a^a)\bar{\tilde{q}}^A
\right|^2 \right\}\,.
\label{pot}
\eeqn
Here  $f^{abc}$ denote the structure constants of the SU$(N)$ group,
$m_A$ is the mass term of the $A$-th flavor,
 and 
the sum over the repeated flavor indices $A$ is implied.

We  introduced the FI $D$-term for the U(1) gauge factor with the FI 
parameter $\xi$. 

Now  we briefly review the vacuum structure and
the excitation  spectrum in  the bulk theory.
The  vacua of the theory (\ref{model}) are determined by the zeros of 
the potential (\ref{pot}). The adjoint fields develop the following
vacuum expectation values (VEVs):
\beq
\left\langle \left(\frac12\, a + T^a\, a^a\right)\right\rangle = - \frac1{\sqrt{2}}
 \left(
\begin{array}{ccc}
m_1 & \ldots & 0 \\
\ldots & \ldots & \ldots\\
0 & \ldots & m_N\\
\end{array}
\right),
\label{avev}
\eeq
For generic values of the quark masses, the  SU$(N)$ subgroup of the gauge 
group is
broken down to U(1)$^{N-1}$. However, in the special limit
\beq
m_1=m_2=...=m_N,
\label{equalmasses}
\eeq
the  SU$(N)\times$U(1) gauge group remains  unbroken by the adjoint field.
In this limit the theory acquires the global flavor SU$(N)$ symmetry.

We can exploit gauge rotations to
make  all squark VEVs real. Then
in the case at hand they take the color-flavor locked form
\beqn
\langle q^{kA}\rangle &=&\sqrt{
\xi}\,
\left(
\begin{array}{ccc}
1 & \ldots & 0 \\
\ldots & \ldots & \ldots\\
0 & \ldots & 1\\
\end{array}
\right),
\qquad \langle \bar{\tilde{q}}^{kA}\rangle =0,
\nonumber\\[4mm]
k&=&1,..., N\qquad A=1,...,N\, ,
\label{qvev}
\eeqn
where we write down the quark fields as an $N\times N$ matrix in 
the color and flavor indices.
This particular form of the squark condensates is dictated by the
third line in Eq.~(\ref{pot}). Note that the squark fields stabilize
at non-vanishing values entirely due to the U(1) factor 
represented by the second term in the third line.

The  vacuum field (\ref{qvev}) results in  the spontaneous
breaking of both gauge and flavor SU($N$) symmetries.
A diagonal global SU$(N)_{C+F}$ survives, however,
\beq
{\rm U}(N)_{\rm gauge}\times {\rm SU}(N)_{\rm flavor}
\to {\rm SU}(N)_{C+F}\,.
\label{c+f}
\eeq
Thus, a color-flavor locking takes place in the vacuum.
The presence of the global SU$(N)_{C+F}$ group is a key reason for the 
formation of non-Abelian strings.
 For generic quark masses the  global symmetry  (\ref{c+f}) is broken down to 
U(1)$^{(N-1)}$.

Let us move on to  the issue of the excitation spectrum  in this vacuum
\cite{VY,ABEKY}.
The mass matrix for the gauge fields $(A^{a}_{\mu},
A_{\mu})$ can be read off from the quark kinetic terms  in (\ref{model}).
It shows that all SU$(N)$ gauge bosons become massive, with
one and the same mass
\beq
m_{W} =g_2\,\sqrt\xi\,.
\label{msuN}
\eeq
The equality of the masses is no accident. It is a consequence of the
unbroken SU$(N)_{C+F}$ symmetry (\ref{c+f}).
The mass of the U(1) gauge boson is
\beq
m_{\gamma} =g_1\,\sqrt{\frac{N}{2}\,\xi}\,.
\label{mu1}
\eeq
Thus, the bulk theory is fully Higgsed.
The mass spectrum of the adjoint scalar excitations is the same as the one 
for the gauge bosons. This is enforced by \ntwo.

The mass spectrum of the quark  excitations can be read off from the potential (\ref{pot}).
We have  $4N^2$ real degrees of freedom of quark scalars $q$ and $\tilde q$.
Out of those 
$N^2$ are eaten up by the Higgs mechanism.
The remaining $3N^2$ states split in three plus $3(N^2-1)$ states with masses
(\ref{mu1}) and (\ref{msuN}), respectively. Combining these states with
the massive gauge bosons and the adjoint scalar states we get  \cite{VY,ABEKY}
one long \ntwo BPS multiplet (eight real bosonic
plus eight fermionic degrees of freedom) with mass (\ref{mu1})
and $N^2-1$ long \ntwo BPS multiplets with mass (\ref{msuN}).
Note that these supermultiplets come in representations of the unbroken
SU$(N)_{C+F}$ group, namely, the singlet and adjoint representations.

Now let us have a closer look at quantum effects in the theory
(\ref{model}). The SU$(N)$ sector is
asymptotically free. The running of the corresponding gauge coupling,
if not interrupted, would drag the theory into the strong coupling regime.
This would invalidate our quasiclassical analysis. Moreover, strong
coupling effects on the Coulomb branch would break SU$(N)$ gauge subgroup
 down to
U(1)$^{N-1}$ by virtue of  the Seiberg--Witten mechanism \cite{SW1}.
No non-Abelian strings would emerge.

The semiclassical analysis above is valid if the FI parameter $\xi$ 
is large,
\beq
\xi\gg \Lambda\, ,
\label{weakcoupling}
\eeq
where $\Lambda$ is the scale of the SU($N$) gauge theory.
This condition ensures weak coupling in the SU$(N)$ sector because
the SU$(N)$ gauge coupling does not run below the scale of the quark VEVs
which is determined by $\xi$. More explicitly,
\beq
\frac{8\pi^2}{g^2_2 (\xi)} =
N\ln{\frac{\sqrt{\xi}}{\Lambda}}\gg 1 \,.
\label{coupling}
\eeq

\section{Towards smaller \boldmath{$\xi$}}
\label{3three}
\label{smallerxi}

Below we will see   that if we pass to small  $\xi$ along the line $\Delta m =0$, into the 
strong coupling domain,  where  the condition (\ref{weakcoupling}) is not met,
the theory undergoes a crossover transition into the Seiberg--Witten Abelian confinement
regime. In this regime the low-energy perturbative sector contains no
nontrivial representations of the unbroken
SU$(N)_{C+F}$ group. Moreover, no non-Abelian strings develop.

The main tool which allows us to identify this crossover transition
is the presence of the global unbroken SU$(N)_{C+F}$ symmetry in the theory at hand.
First, we note that it is not spontaneously broken in the bulk. If it
were broken this would imply the presence of massless Goldstone states.
However, we showed above that the perturbative sector of the theory
at large $\xi$ has a large mass gap of the order of $g^2\sqrt{\xi}$, and 
masses of no  states can be shifted to zero by small quantum corrections of the order of 
$\Lambda$. Nor do we see the adjoint multiplet of massless Goldstones at
small $\xi$.

The presence of the global unbroken  SU$(N)_{C+F}$ symmetry means that all
multiplets should come in  representations of this group. We showed
above that  at large $\xi$ this is the case indeed: all light states come
in adjoint and  singlet representations ($(N^2-1)+1$). We will see
later that at small $\xi$ (in the Seiberg--Witten regime) the low-energy  
spectrum is very different. It contains only $N$ states which do not
fill any nontrivial representations of  SU$(N)_{C+F}$. They are all
singlets.

To elucidate the point let us note the following.
All $(N^2-1)$ states of the adjoint gauge-boson multiplet
of SU$(N)_{C+F}$ have degenerate masses (\ref{msuN}) at large $\xi$.
The presence of the unbroken global SU$(N)_{C+F}$ ensures that they are not
split. Imagine that these states were split with small splittings
of the order of $\Lambda$. Then in the limit of small $\xi$, $\xi \ll \Lambda$,
some of these states could, in principle,  evolve into $(N-1)$ light Abelian states while other
members of the multiplet could acquire large masses, of the order of $\Lambda$ 
(i.e. the photons could become  light, while the $W$ bosons could become heavy).
We stress that this does not happen in the theory at hand. The adjoint
multiplet is not split at large $\xi$ and therefore can
disappear from the low-energy spectrum at small $\xi$ only as a whole.
Hence, the light photons in the Seiberg--Witten regime at small $\xi$ have nothing to 
do with the diagonal (Cartan) entries of the gauge adjoint  SU$(N)_{C+F}$ multiplet at large
$\xi$. Similarly, the light dyons in the Seiberg--Witten regime 
at small $\xi$ have nothing 
to do with the light quarks $q^{kA}$, $\tilde{q}_{Ak}$ of the non-Abelian confinement
regime at
large $\xi$. The latter fill  the adjoint representation of  SU$(N)_{C+F}$ 
(see the  discussion above), while the former are singlets.

In order to see what happens with the low-energy spectrum of the bulk theory as we 
reduce $\xi$ we use the following method. First we introduce quark mass
differences $(m_A-m_B)$ and take them large, $|m_A-m_B|\gg \Lambda$. Then
we can reduce the parameter $\xi$ keeping the theory at weak coupling and 
under control. Next, we reach the Coulomb branch at zero $\xi$ and
use the exact  solution of the theory \cite{SW1,SW2}
to go back to the desired limit
of degenerate quark masses (\ref{equalmasses}).
Thus, our routing is:  Domain I $\to$ Domain II $\to$ Domain III. 
In domain II the global SU$(N)_{C+F}$ is lost, and a level crossing occurs.

The program outlined above 
will be carried out in full  in Sect.~\ref{coulombbranch}.

To conclude this section we briefly review the theory ({\ref{model})
at non-zero quark mass differences $(m_A-m_B)\neq 0$, see \cite{SYmon,SYrev}.
At non-vanishing $(m_A-m_B)$ the global SU$(N)_{C+F}$ is 
explicitly broken down to
U(1)$^{(N-1)}$. The adjoint multiplet is split. The diagonal entries 
(photons and their \ntwo quark superpartners) have
masses given in (\ref{msuN}), while the off-diagonal states ($W$ bosons and the
off-diagonal entries of the quark matrix  $q^{kA}$)
 acquire additional contributions to their masses 
proportional to $(m_A-m_B)$.  As we make the mass differences larger, the
$W$ bosons become exceedingly  heavier,  decouple from
the low-energy spectrum,  and we are left with $N$ photon states and
 $N$ diagonal elements of the quark matrix. The low-energy spectrum
becomes Abelian.

\section {Non-Abelian strings at large \boldmath{$\xi$}}
\label{strings}
\setcounter{equation}{0}

Here we will study the passage from Domain I $\to$ Domain II.
At first, we will briefly review
non-Abelian strings \cite{HT1,ABEKY,SYmon,HT2} in the theory (\ref{model}),
see  \cite{SYrev} for details. The Abelian $Z_N$-string solutions
 break the  SU$(N)_{C+F}$ global group. Therefore
strings have orientational zero modes, associated with rotations of their color 
flux inside the non-Abelian SU($N$). This makes these strings non-Abelian.
The global group is broken by the $Z_N$ string solution down to 
${\rm SU}(N-1)\times {\rm U}(1)$.
Therefore
the moduli space of the non-Abelian string is described by the coset
\beq
\frac{{\rm SU}(N)}{{\rm SU}(N-1)\times {\rm U}(1)}\sim {\rm CP}(N-1)\,.
\label{modulispace}
\eeq
The CP$(N-1)$ space can be parametrized by
 a complex vector $n^l$
 in the fundamental representation of SU($N$) subject to the constraint
\beq
 n^*_l n^l =1\,,
\label{unitvec}
\eeq
where $l=1, ..., N$.
As we will show below, one U(1) phase will be gauged away in the effective
sigma model. This gives the correct number of degrees of freedom,
namely, $2(N-1)$.

With this parametrization the elementary string solution 
(with the lowest winding number) can be
written as \cite{SYmon,GSY05}
\beqn
q &=& \frac1N[(N-1)\phi_2 +\phi_1] +(\phi_1-\phi_2)\left(
n\,\cdot n^*-\frac1N\right) ,
\nonumber\\[3mm]
A^{{\rm SU}(N)}_i &=& \left( n\,\cdot n^*-\frac{1}{N}\right)
\varepsilon_{ij}\, \frac{x_i}{r^2}
\,
f_{NA}(r) \,,
\nonumber\\[3mm]
A^{{\rm U}(1)}_i &=& \frac1N
\varepsilon_{ij}\, \frac{x_i}{r^2} \, f(r) \, , \qquad  
\bar{\tilde{q}}^{kA} =0,
\label{str}
\eeqn
where $i=1,2$ labels coordinates in the plane orthogonal to the string
axis and $r$ and $\alpha$ are the polar coordinates in this plane.
For brevity we suppress all SU$(N)$  indices. The profile
functions $\phi_1(r)$ and  $\phi_2(r)$ determine the profiles of
the scalar fields,
while $f_{NA}(r)$ and $f(r)$ determine the SU($N$) and U(1)
gauge fields of the
string solution, respectively. These functions satisfy the
first-order equations \cite{ABEKY} which can be solved numerically.

The tension of the elementary string is given by
\beq
T \, = \, 2\pi\,\xi.
\label{ten}
\eeq

Making the moduli vector $n^l$ a slowly varying function of the string world
sheet coordinates $x_k$ ($k=0,3$), we can derive 
the  effective low energy-theory on the string world sheet 
\cite{ABEKY,SYmon,GSY05}. From the topological 
reasoning above (see (\ref{modulispace}))   it is clear that we will get
two-dimensional CP$(N-1)$ model.
The \ntwo supersymmetric CP$(N-1)$ model    can be understood as
a strong-coupling limit  of a U(1) gauge theory \cite{W93}. Then the  bosonic part
of the  action takes the form
\beqn
S_{{\rm CP}(N-1)}
& =&
\int d^2 x \left\{
 2\beta\,|\nabla_{k} n^{\ell}|^2 +\frac1{4e^2}F^2_{kl} + \frac1{e^2}
|\pt_k\sigma|^2
\right.
\nonumber\\[3mm]
 &+&    4\beta\,|\sigma|^2 |n^{\ell}|^2 + 2e^2 \beta^2(|n^{\ell}|^2 -1)^2
\Big\}\,,
\label{cpg}
\eeqn
where $\nabla_k= \partial_k - i A_k $ while $\sigma$ is a complex scalar
field. The condition (\ref{unitvec}) is
implemented in the limit $e^2\to\infty$. Moreover, in this limit
the gauge field $A_k$  and its \ntwo bosonic superpartner $\sigma$ become
auxiliary and can be eliminated by virtue of the equations of motion,
\beq
A_k =-\frac{i}{2}\, n^*_\ell \stackrel{\leftrightarrow}
{\partial_k} n^\ell \,,\qquad \sigma=0\,.
\label{aandsigma}
\eeq

The two-dimensional coupling constant $\beta$ here is determined by the
four-dimensional non-Abelian coupling via the relation
\beq
\beta= \frac{2\pi}{g_2^2}\,.
\label{betag}
\eeq
The above relation between the four-dimensional and two-dimensional coupling
constants (\ref{betag}) is obtained  at the classical level 
\cite{ABEKY,SYmon}.
 In quantum theory
both couplings run. In particular, the CP $(N-1)$ model is asymptotically free
\cite{Po3} and develops its own scale $\Lambda_{\sigma}$. 
The ultraviolet cut-off of the sigma model on the string worldsheet
is determined by  $g_2\sqrt{\xi}$. 
Equation~(\ref{betag}) relating the two- and four-dimensional couplings
is valid at this scale, implying
\beq
\Lambda^{N}_{\sigma} = g_2^N\xi^{\frac{N}{2}}
 e^{-\frac{8\pi^2}{g_2^2}} =\Lambda^N\, .
\label{lambdasig}
\eeq

Note that in the bulk theory {\em per se}, because of the VEVs of
the squark fields, the coupling constant is frozen at
$g_2\sqrt{\xi}$; there are no logarithms below this scale.
The logarithms of the string worldsheet theory take over.
Moreover, the dynamical scales of the bulk and worldsheet
theories turn out to be the same \cite{SYmon}.

The CP$(N-1)$ model was solved by Witten
in the large-$N$ limit \cite{W79}.
We will briefly summarize Witten's results and translate them in terms
of strings in four dimensions \cite{SYmon}.

Classically the  field $n^{\ell}$ can have arbitrary
direction; therefore, one might naively expect
a spontaneous breaking of SU($N$) and
the occurrence of massless Goldstone modes on the string world sheet.
Well, this cannot happen
in two dimensions. Quantum effects restore the
symmetry. Moreover, the condition
(\ref{unitvec}) gets in effect relaxed. Due to strong coupling
we have more degrees of freedom than in the original Lagrangian,
namely all $N$ fields $n$ become dynamical and acquire
masses $ \Lambda_{\sigma}$.

Deep in the quantum non-Abelian regime the
CP$(N-1)$-model strings
carry no average SU($N$) magnetic flux.
To see that this is indeed the case,
note that the SU($N$) magnetic flux of the non-Abelian string
(\ref{str}) is given by
\beq
\int d^2 x \, (F^{*}_3)_{\,\,\,{\rm SU}(N)}=2\pi 
\left( n\,\cdot n^*-\frac{1}{N}\right)\,,
\label{strflux}
\eeq
where 
\beq
F^{*}_i=\mbox{$\frac{1}{2}$}\,\varepsilon_{ijk}F_{jk}\,, \quad (i,j,k=1,2,3).
\label{ccff}
\eeq
As was shown by Witten  \cite{W79},  in the CP$(N-1)$ model strong quantum fluctuations 
of 
$n^\ell$ result in
\beq
\langle n^\ell \rangle =0\,,
\label{zeronl}
\eeq
implying, in turn, that the average SU($N$) magnetic flux of the non-Abelian string 
vanishes. 
We will use this circumstance later, to distinguish between large-$\xi$
non-Abelian   and small-$\xi$  Abelian $Z_N$ strings in the Seiberg--Witten regime
below the crossover.
The latter do carry the magnetic flux  directed inside the Cartan subalgebra of SU($N$).
The CP$(N-1)$ model has $N$ vacua \cite{W79}.
They are interpreted in the problem at hand as $N$ different elementary non-Abelian strings.
These $N$ vacua differ from each other by the expectation value
of the chiral bifermion operator, see e.g. \cite{NSVZsigma}.
At strong coupling  the chiral condensate is the
order parameter for $Z_N$ breaking (instead of the flux, see Appendix). The U(1) chiral 
symmetry of 
the CP$(N-1)$ model is 
explicitly broken to a discrete $Z_{2N}$ symmetry by the chiral anomaly
(for a discussion of the global symmetry on the string world sheet
see Appendix).
The bifermion condensate  breaks $Z_{2N}$ down to $Z_2$. That's the
origin of the $N$-fold degeneracy of the vacuum state.

Now, to make our consideration simpler   we will focus on the simplest case
$N=2$. For arbitrary $N$ the emerging dynamical pattern is similar. 
The solution for the non-Abelian
string (\ref{str}) in the $N=2$ case takes the form
\begin{eqnarray}
&&
q\,=\, \frac12 (\phi_1+\phi_2)
+\frac{\tau^a}{2}S^a (\phi_1-\phi_2) ,\qquad \tilde{q}=0,
\nonumber \\[4mm]
&&
A^a_{i}(x)
=
S^a \,\varepsilon_{ij}\, \frac{x_j}{r^2}\,
f_{NA}(r)\, ,\qquad
A_{i}(x)
= \varepsilon_{ij} \, \frac{x_j}{r^2}\,
f(r)\, ,
\label{sna}
\end{eqnarray}
where 
$ S^a$ ($a=1,2,3$) is a real moduli vector
subject to the constraint
\beq
 (S^{a})^2 = 1\,.
\label{ensq}
\eeq
Its relation to the complex vector $n^\ell$ is as follows
\beq
S^a=\bar{n}\,\tau^a n.
\label{Sn}
\eeq
We have CP(1) as the
effective world-sheet theory. It is equivalent to the
O(3) sigma model. In terms of real vector $S^a$ the bosonic part of world-sheet theory
 has the form
\beq
S=\beta\,\int d^2 x \,\frac12\left(\pt_k S^a\right)^2\,.
\label{o3}
\eeq

Now let us introduce quark mass differences $(m_A-m_B)$. In the   $N=2$ case
we have just one (generally speaking complex) parameter 
\beq
\Delta m=m_1-m_2\,.
\eeq
The vacuum expectation values of the adjoint field reduce to
\beq
\langle a^3 \rangle = -\,\frac{\Delta m}{\sqrt{2}}, \qquad
\langle a \rangle =-\sqrt{2}\;\frac{m_1+m_2}{2},
\label{N2avev}
\eeq
see (\ref{avev}). The non-Abelian string (\ref{sna}) is no longer a
solution of the first-order equations for arbitrary $S^a$. 
The global
SU(2)$_{C+F}$ is explicitly broken down to U(1) by  $\Delta m\neq 0$. Nevertheless,
if we keep $\Delta m$ small, we can consider $S^a$ as   quasimoduli,
with a shallow potential on the  CP(1)  moduli space. The string
solution (\ref{sna}) in this case should be supplemented by a
nontrivial profile for the adjoint field \cite{SYmon,SYrev},
\beq
a^a=-\frac{\Delta m}{\sqrt{2}}\, \left[\delta^{a3}\, \frac{\phi_1}{\phi_2} 
 +S^a\,  S^3\,
(1-\frac{\phi_1}{\phi_2} )\right]\, .
\label{aa}
\eeq
Plugging this modified string solution in the action of the theory gives 
\cite{SYmon,SYrev}
the effective string world-sheet theory:
\ntwo CP(1) model with  twisted mass \cite{Alvarez}. The bosonic part of
the action is
\beq
\label{o3mass}
S_{{\rm CP}(1)}=\beta \int d^2 x \left\{\frac12 \left(\pt_k
S^a\right)^2 +\frac{|\Delta m|^2}{2}\,\left( 1-S_3^2\right)\right\}\, .
\eeq
This is the only functional form that allows \ntwo completion.
The mass-splitting parameter $\Delta m$ of the bulk
theory exactly coincides with the twisted mass of the world-sheet  model.

The CP(1) model (\ref{o3mass}) has two vacua located at 
$S^a=(0,0,\pm 1)$. Clearly these two vacua correspond
to two elementary  $Z_2$ strings.

With non-vanishing $\Delta m$ we can introduce a gauge invariant quantity which
measures the SU$(2)$ non-Abelian flux of the string. We define\,\footnote{The 
subscript 3 indicates the direction along the string axis, cf. Eq.~(\ref{strflux}).}
\beq
\Phi=\int d^2 x \, a^a F^{*a}_3.
\label{flux}
\eeq
This order parameter will be used  below to distinguish between different regimes
 of the theory.

Substituting (\ref{aa}) and (\ref{strflux}) into (\ref{flux}) we 
get
\beq
\Phi=-2\pi\,\frac{\Delta m}{\sqrt{2}}\, S^3 
\label{fluxna}
\eeq
At small $\Delta m$, $\Delta m\ll\Lambda_{\sigma}$,
 the fields $S^a$ strongly fluctuate and
$\langle S^3 \rangle=0$ (see (\ref{zeronl})). Therefore, 
\beq
\langle\Phi  \rangle _{\rm I} \to 0\,\,\,{\rm at}\,\,\, \Delta m \ll \Lambda_{\sigma}\,,
\label{fluxI}
\eeq
where the subscript I refers to the non-Abelian domain at large $\xi$
and small $\Delta m$, as it is indicated  in Fig.~\ref{figphasediag}. 

Instead, at large $\Delta m$ ($\Delta m\gg \Lambda_{\sigma}$) the O(3) sigma
 model (\ref{o3mass}) is at weak coupling. Fluctuations
are small, and the $S^a$ acquires vacuum values at the north and
south poles of the $S_2$ sphere, $\langle S^a\rangle =(0,0,\pm 1)$.
As a result
\beq
\langle\Phi  \rangle _{\rm II} \to \mp 2\pi\,\frac{\Delta m}{\sqrt{2}}\,,
\label{fluxII}
\eeq
where the subscript II marks domain II in Fig.~\ref{figphasediag}.

We see that the behavior of the string flux $\Phi$ drastically changes
as we pass from the non-Abelian domain I of  large $\xi$
and small $\Delta m$ to the Abelian domain II
of large $\Delta m$,
see Fig.~\ref{figflux}. At large $\Delta m$ this theory is in the weak coupling
 regime and 
 fluctuations are small, $\langle S^3  \rangle \approx \pm 1$ and 
the flux $\Phi$ is given by (\ref{fluxII}). At small $\Delta m$
the world sheet theory is in the strong coupled quantum regime, fluctuations are large
and the vector $S^a$ is smeared over the whole sphere. Therefore,
$\langle S^3  \rangle\approx 0$ and $\langle\Phi \rangle\approx 0$. 
The crossover  between these
two regimes is at $\Delta m \sim \Lambda_{\sigma}$.  Note, that the drastic
change of behavior in
the world-sheet CP(1)
model is correlated with the dynamics of the bulk 
theory.

\begin{figure}
\epsfxsize=7cm
\centerline{\epsfbox{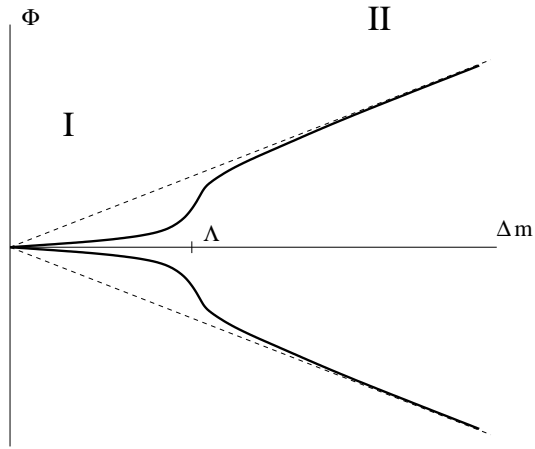}}
\caption{\small Flux (\ref{flux}) as a function of $\Delta m$ in domains
 I and II.}
\label{figflux}
\end{figure}

In Sect.~\ref{bulk} we saw
that the perturbative spectrum of the bulk theory is different in 
these two domains: it is essentially non-Abelian in the domain I while
in the domain II the $W$ bosons become exceedingly heavier and decouple from
the low-energy spectrum: we  are left with $N$ photons and their superpartners,
the diagonal elements of the quark matrix. The pattern repeats itself at
the nonperturbative level: the non-Abelian strings
evolve into the Abelian strings as we increase $|\Delta m |$.

Later we will see  that the crossover becomes exceedingly more pronounced as we 
increase the number of colors $N$. In the limit $N\to\infty$ the crossover evolves 
 into a genuine phase 
transition. Note also that in nonsupersymmet\-ric theories we do have
a phase transition between the phase with non-Abelian strings at small
 $|m_A-m_B|$ 
and the phase with the Abelian strings at large  $|m_A-m_B|$  \cite{GSYphtr}.
It is related to the restoration of the broken discrete $Z_N$ symmetry at 
small  $ |m_A-m_B|$.\footnote{At $N>2$ the above discrete symmetry  of the Lagrangian
takes place under a special choice of the mass parameters, see Appendix.}
 In the supersymmetric theory at hand the discrete $Z_N$
symmetry is always broken (by VEVs of $n^\ell$ at large  $ |m_A-m_B|$  or
by the two-dimensional bifermion condensates at small  $ |m_A-m_B|$, see Appendix).
 Therefore
in the supersymmetric case we have a crossover rather than a phase transition.
 
To conclude this section, we briefly review the world-sheet theory on 
the non-Abelian string for generic $N$. It is described by
the twisted-mass-deformed CP$(N-1)$ model. It can be nicely written \cite{Dorey} as a
strong coupling limit of a U(1) gauge theory.  
With twisted masses of the $n^{\ell}$ fields taken into account,  
the bosonic part of the action
(\ref{cpg}) takes the form
\beqn
S 
&=&
\int d^2 x \left\{
 2\beta\,|\nabla_{k} n^{\ell}|^2 +\frac1{4e^2}F^2_{kl} + \frac1{e^2}
|\pt_k\sigma|^2
\right.
\nonumber\\[3mm]
 &+& \left. 4\beta\,\left|\sigma-\frac{\tilde{m}_{\ell}}{\sqrt{2}}\right|^2 
|n^{\ell}|^2 +
 2e^2 \beta^2(|n^{\ell}|^2 -1)^2
\right\}\,.
\label{mcpg}
\eeqn
where
\beq
\tilde{m}_{\ell}=m_{\ell}-m,\;\;\; m\equiv \frac1N \sum_{\ell}m_{\ell}\,,
\label{tildem}
\eeq
and the sum over $\ell$ in (\ref{mcpg}) is implied.

\section {The theory in domain III}
\label{coulombbranch}
\setcounter{equation}{0}

Now, 
we will consider the passage from domain II to domain III.
In order to study the theory in the regime III (see Fig.~\ref{figphasediag})
we first assume the quark mass differences to be large. Then the theory stays at
weak coupling and we can safely decrease the value of the FI parameter $\xi$. Next,
we use the exact Seiberg--Witten solution of the theory on the Coulomb branch
(at   $\xi=0$) to pass from regime II to regime III. To simplify our discussion,
we will consider here only the case $N=2$.

\subsection{The \boldmath{$r=2$} quark vacuum}
\label{tr2qv}

Our first task is to identify the $r=2$ quark vacuum
(which we described semiclassically above) using the exact
Seiberg--Witten solution \cite{SW1,SW2} \footnote{These solutions were obtained by
Seiberg and Witten in the SU(2) gauge theories. Generalizations
to SU$(N)$ were obtained in \cite{ArFa,KLTY,ArPlSh,HaOz}.}.
The Seiberg--Witten curve for the U(2) gauge theory with $N_f=2$  flavors
has the form 
\beq
y^2= (x-\phi_1)^2 (x-\phi_2)^2  - 
 4\Lambda^2\, \left(x+\frac{m_1}{\sqrt{2}}\right)\left(x+\frac{m_2}{\sqrt{2}}\right),
\label{U2curve}
\eeq
where $\phi_1$ and $\phi_2$ are gauge-invariant parameters on the Coulomb 
branch.

Semiclassically
\beq
\phi_1\approx a_1\equiv\frac12(a+a^3), \qquad \phi_2\approx a_2\equiv\frac12(a-a^3).
\label{classphia}
\eeq
Let us make a shift in the variable $x$ introducing a new variable $z$,
\beq
x=-\frac{m}{\sqrt{2}} +z , \qquad m=\frac12(m_1+m_2).
\label{z}
\eeq

With $\Delta m\gg \Lambda $ we identify the $r=2$ singularity, the point where
both  quarks $q^{11}$ and $q^{22}$ become massless. Upon switching on $\xi
\neq 0$, this $r=2$
singularity turns into the $r=2$ vacuum we considered in the semiclassical
approximation in the previous sections.

It turns out  that in the $r=2$
vacuum
\beq
\phi_1 +\phi_2=-2\frac{m}{\sqrt{2}}\,.
\eeq
We parametrize the deviations of $\phi_1$ and $\phi_2$
from their mean value $-\frac{m}{\sqrt{2}}$ by a new parameter 
$\phi$,
\beqn
\phi_1&=& -\frac{m}{\sqrt{2}}+\phi\,, 
\nonumber\\[3mm]
\phi_2 &=&
-\frac{m}{\sqrt{2}} -\phi\,.
\label{bphi}
\eeqn
With this parametrization the curve (\ref{U2curve}) reduces to\,\footnote{ $\Delta 
m^2$ is a shorthand for 
$(\Delta m)^2$.} 
\beqn
y^2
&=&
 (z-\phi)^2 (z+\phi)^2  - 
 4\Lambda^2 \, \left(z+\frac{\Delta m}{2\sqrt{2}}\right)
\left(z-\frac{\Delta m}{2\sqrt{2}}\right)
\nonumber\\[3mm]
&=&
(z^2-\phi^2)^2   - 
 4\Lambda^2\, \left(z^2-\frac{\Delta m^2}{8}\right).
\label{U2redcurve}
\eeqn

Next, we look for the values of the parameter $\phi$ which ensure that this curve
has two double roots associated with two quarks being massless. 
This curve is a perfect square
\beq
y^2=\left[z^2-\frac14 \left(\frac{\Delta m^2}{2} +4\Lambda^2\right)\right]^2,
\label{sqcurve}
\eeq
see \cite{Dorey}, at
\beq
\phi= \frac{1}{2}
\sqrt{\frac{\Delta m^2}{2}- 4\Lambda^2}\,\,.
\label{b}
\eeq
In fact,
there are two solutions with plus and minus signs in front of the 
square root above.  They correspond to $\phi_1$ and $\phi_2$, 
namely
\beqn
\phi_1
&=&
-\frac{m}{\sqrt{2}} -\frac12\sqrt{\frac{\Delta m^2}{2}-4\Lambda^2}\,\,, 
\nonumber\\[3mm]
\phi_2
&=&
-\frac{m}{\sqrt{2}}+\frac12\sqrt{\frac{\Delta m^2}{2}-4\Lambda^2}\,\,.
\label{U2phivev}
\eeqn
In the semiclassical limit $\Delta m\gg \Lambda$ these solutions
reduce to
\beq
\phi_1\approx -\frac{m_1}{\sqrt{2}}\, , \qquad \phi_2\approx 
-\frac{m_2}{\sqrt{2}}\,,
\label{classphi}
\eeq
which coincides with Eq.~(\ref{avev}). This means that we  correctly identified 
the $r=2$ 
quark vacuum where two quarks $q^{11}$ and $q^{22}$
are massless, see Sect.~\ref{bulk}.

Two double  roots of the curve in the quark  vacuum are 
\beqn
e_1
&=&
e_2
=
-\frac{m}{\sqrt{2}} -\frac12\sqrt{\frac{\Delta m^2}{2}+4\Lambda^2}\,\,,
\nonumber\\[3mm]
e_3
&=&
e_4
=
-\frac{m}{\sqrt{2}} +\frac12\sqrt{\frac{\Delta m^2}{2}+4\Lambda^2}\,\,.
\label{U2roots}
\eeqn

The Seiberg--Witten exact solution of the theory relates 
VEVs of the fields $a$, $a^3$ and $a_D$, $a_D^3$ (which, in turn,
determine the spectrum of the BPS states on the Coulomb branch)
to certain  integrals along $\alpha$ and $\beta$ contours
in the $x$-plane \cite{SW1,SW2}.
Say, the derivatives of VEVs of $a_1$ and  $a_2$ are given by
the following
integrals along the $\alpha$ contours:

\beqn
\frac{\pt a_1}{\pt \phi_1} & = & \frac{1}
{2\pi i}\,
\int_{\alpha_1}dx\frac{x-\phi_2}{y}\,,
\nonumber\\[3mm]
\frac{\pt a_1}{\pt \phi_2} & = & \frac{1}
{2\pi i}\,
\int_{\alpha_1}dx\frac{x-\phi_1}{y}\,,
\nonumber\\[3mm]
\frac{\pt a_2}{\pt \phi_1} & = & \frac{1}
{2\pi i}\,
\int_{\alpha_2}dx\frac{x-\phi_2}{y}\,,
\nonumber\\[3mm]
\frac{\pt a_2}{\pt \phi_2} & = & \frac{1}
{2\pi i}\,
\int_{\alpha_2}dx\frac{x-\phi_1}{y}\,,
\label{aintegrals}
\eeqn
while the derivatives of $a_D$'s are given by similar integrals
along the $\beta$ contours.

The presence of 
two massless quarks $q^{11}$ and $q^{22}$
in the $r=2$ vacuum at $\Delta m\gg\Lambda$
implies
\beq
a_1+\frac{m_1}{\sqrt{2}}=0,\qquad a_2+\frac{m_2}{\sqrt{2}}=0\,.
\label{masslessquarks}
\eeq
Thus, the fields $a_1$, $a_2$ are regular at the singularity while the
fields $a_D$ have logarithmic divergences related to the $\beta$
functions of the low-energy U(1)$\times$U(1) theory. This
ensures that the $\alpha_1$ contour should go around the
roots $e_1$, $e_2$ while the $\alpha_2$ contour should go around the
roots $e_3$, $e_4$. In the $r=2$ vacuum (\ref{U2roots})
both contours shrink and produce regular $a$'s.
 The basis of the $\alpha$ and $\beta$ contours
is shown in Fig.~\ref{figcontours}. Here we consider our U(2) theory
as a two-flavor SU(3) gauge theory  broken down to
U(2) at a very high scale. In terms of the SEiberg--Witten curve this corresponds
to extra two roots of the SU(3) curve being far away from
four roots of the U(2) curve (\ref{U2curve}).

\begin{figure}
\epsfxsize=8.5cm
\centerline{\epsfbox{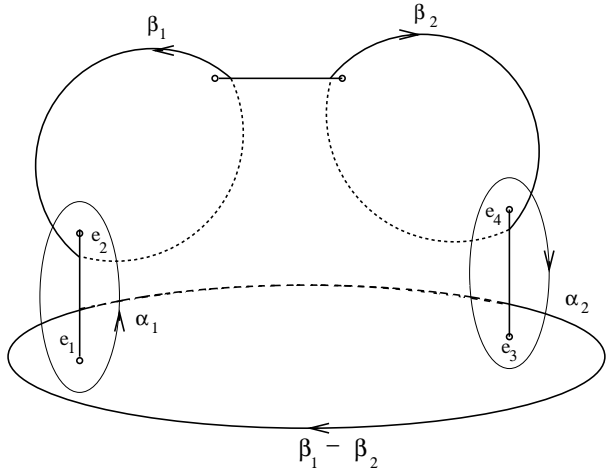}}
\caption{\small Basis of the  $\alpha$ and  $\beta$ contours of our U(2) gauge theory
viewed as an SU(3) theory broken down to U(2). Two extra roots of the SU(3) theory are
far away in the $x$ plane.}
\label{figcontours}
\end{figure}

As a double check of our identification of the quark vacuum let us  calculate 
the derivatives (\ref{aintegrals}) in the semiclassical
approximation $\Delta m\gg\Lambda$. Substituting (\ref{sqcurve})
into (\ref{aintegrals}) we get at $\Delta m\gg\Lambda$
\beqn
\frac{\pt a_1}{\pt \phi_1} & \approx & 1\,,\qquad
\frac{\pt a_1}{\pt \phi_2}  \approx  0\,,
\nonumber\\[3mm]
\frac{\pt a_2}{\pt \phi_1} & \approx & 0\,,
\qquad
\frac{\pt a_2}{\pt \phi_2} \approx  1\,.
\label{aphi}
\eeqn
This is in accord  with (\ref{classphia}) and confirms
our choice of the $\alpha$ contours in Fig.~\ref{figcontours}.

To conclude this subsection we note that the monopole singularity
(the point on the Coulomb branch where the SU(2) monopole becomes
massless, $a_D^3=0$) corresponds to shrinking of the $(\beta_1-\beta_2)$ contour, 
i.e, in other words, to $e_1=e_3$.

\subsection{Monodromies}
\label{monodromies}

Let us study how the quantum numbers of massless quarks $q^{11}$ and $q^{22}$
change as we reduce $|\Delta m|$ and go from domain II into domain 
III where the theory is at strong coupling. The quantum numbers change  due to monodromies 
with respect to $\Delta m$. The 
complex plane of $\Delta m$ has cuts and when we cross these cuts, the $a$ and $a_D$
fields acquire monodromies and the quantum numbers of states change accordingly.
Monodromies with respect to quark masses were studied in \cite{BF} in the theory
with the SU(2) gauge group using a monodromy matrix approach.

Here we will investigate the monodromies in the U(2) theory with two quark flavors
using a slightly different approach, similar to that of Ref.~\cite{CKM}. If two roots of the Seiberg--Witten curve coincide, the contour
which goes around these roots shrinks and produces a regular potential. Say, as was
discussed above, at $\Delta m\gg \Lambda$ we have two double roots $e_1=e_2$
and $e_3=e_4$ in the $r=2$ vacuum. Thus, two contours $\alpha_1$ and $\alpha_2$ 
shrink (see Fig.~\ref{figcontours}), and potentials $a_1$ and $a_2$ are regular. This is 
 associated with masslessness of two quarks, see (\ref{masslessquarks}).

Instead, in the monopole singularity $e_1=e_3$; thus 
the $(\beta_1-\beta_2)$ contour
shrinks producing a regular $a_D^3$. This is associated with the masslessness
of the SU(2) monopole, $a_D^3=0$ \cite{SW1}.

If we decrease $|\Delta m|$ and cross the cuts in the $\Delta m$ plane, the root pairing
 in the given vacuum may change. This would mean that a different
combination of $a$ and $a_D$ becomes regular implying a change of 
the quantum numbers
of the massless states in the  given vacuum. To see how it works for our $r=2$ vacuum
we go to the Argyres--Douglas (AD) point point\,\cite{AD,APSW}.
The AD point is a 
particular value of the quark mass parameters where more mutually nonlocal states
become massless. In fact, we will study the collision of the
$r=2$ quark vacuum
with the monopole singularity. We approach the AD point from domain II
at large $\Delta m$. We will show below that as we pass through the AD point
the root pairings   change in the $r=2$ vacuum implying a change of the 
quantum numbers of the massless states. Two massless quarks transform into two
massless dyons. 

To be more precise, we collide the $r=2$ vacuum with two massless quarks with 
the quantum 
numbers
\beqn
&&(n_e,n_m;n_e^3,n_m^3)=(1/2,0;1/2,0), \nonumber\\[2mm]
&&(n_e,n_m;n_e^3,n_m^3)=(1/2,0;-1/2,0)
\label{quarks}
\eeqn
 with the monopole singularity with
\beq
(n_e,n_m;n_e^3,n_m^3)=(0,0;0,1)
\label{monopole}
\eeq
 where the monopole becomes massless. Here $n_e$ and $n_m$ denote electric and 
magnetic charges of a state with respect to U(1) gauge group, while $n_e^3$ and 
$n_m^3$ stands for electric and magnetic charges  with respect to the Cartan
generator of SU(2) gauge group (broken down to U(1) by $\Delta m$).

As was already mentioned, the $(0,0;\,0,1)$ monopole is massless if $e_1=e_3$. 
Equation 
(\ref{U2roots}) shows that this can happen in the $r=2$ vacuum only if all four
roots of the U(2) curve coincide at 
\beq
\Delta m^2=-8\Lambda^2, \qquad e_1=e_2=e_3=e_4=-\frac{m}{\sqrt{2}}\,.
\label{ADpoint}
\eeq
This is the  position of the AD point where both, the quarks and 
the SU(2) monopole become simultaneously massless.

In order to see how the root pairings   in the $r=2$ vacuum change as we decrease 
$|\Delta m|$ and pass from domain II into domain III through the AD point
(\ref{ADpoint}) we have to slightly split the roots by shifting $\phi$ 
from its $r=2$ solution (\ref{b}). Let us take
\beq
\phi^2= \frac{1}{4}\left(
\frac{\Delta m^2}{2}- 4\Lambda^2\right) +\frac1{4\Lambda^2}\,\delta^2\,,
\label{delta}
\eeq
where $\delta$ is a small deviation.
Then the curve (\ref{U2curve}) can be approximately (at small $\delta$ )
written as 
\beq
y^2\approx \left[z^2-\frac14 \left(\frac{\Delta m^2}{2} +4\Lambda^2\right)\right]^2
-\delta^2.
\label{curvedelta}
\eeq
Now all four roots split as follows:
\beqn
e_1 & = & -\frac{m}{\sqrt{2}}+\sqrt{\mu^2+\delta}\,, \qquad 
e_2=-\frac{m}{\sqrt{2}}+\sqrt{\mu^2-\delta},
\nonumber\\[3mm]
e_3 & = & -\frac{m}{\sqrt{2}} -\sqrt{\mu^2+\delta}\,, \qquad
 e_4=-\frac{m}{\sqrt{2}}-\sqrt{\mu^2-\delta},
\label{splitroots}
\eeqn
where we introduced a shorthand notation
\beq
\mu\equiv \frac12\sqrt{\frac{\Delta m^2}{2}+4\Lambda^2}\,.
\label{mu}
\eeq
This parameter vanishes at the AD point.

In order to pass through the AD point from domain II into domain III we decrease
$|\Delta m|$ keeping $\Delta m$ pure
imaginary,
\beq
\Delta m=|\Delta m|\,e^{i\frac{\pi}{2}}.
\eeq
Then $\mu$ goes along the imaginary axis towards the origin (which is the AD point)
and below the AD point increases along the positive axis.
We also fix the parameter $\delta$ to be imaginary too,
 $\delta =|\delta |\,e^{i\frac{\pi}{2}}$. This is convenient as all four roots stay
split at any $|\Delta m|$.

As we decrease $|\Delta m|$ the roots (\ref{splitroots}) move as shown in 
Fig.~\ref{figroots}. We see that the root pairings in the $r=2$ vacuum change. Namely,
at large $|\Delta m|$ we have (at $\delta=0$)
\beq
e_1=e_2, \qquad e_3=e_4,
\label{IIroots}
\eeq
 which, as was explained above, corresponds to shrinking of the
$\alpha_1$ and $\alpha_2$ contours and masslessness of two quarks
(\ref{quarks}).
\begin{figure}
\epsfxsize=7cm
\centerline{\epsfbox{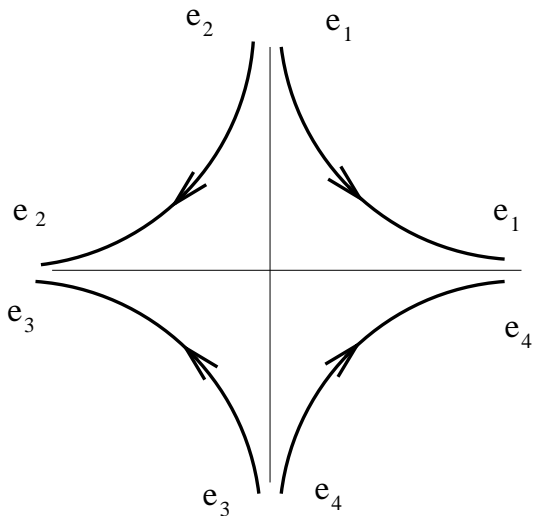}}
\caption{\small As we decrease $|\Delta m|$ (keeping
$\Delta m$ imaginary) and pass through the
AD point, the roots $e_{1,2,3,4}$  move in the $x$ plane.}
\label{figroots}
\end{figure}
Below the AD point at small $|\Delta m|$ we have 
\beq
e_2=e_3, \qquad e_1=e_4,
\label{IIIroots}
\eeq
which corresponds to shrinking of the contours
\beq
\beta_1-\beta_2+\alpha_1\to 0,\qquad -\beta_1+\beta_2+\alpha_2\to 0\,.
\label{shrcont}
\eeq
This means that massless quarks  in the $r=2$ vacuum transformed into
massless  dyons $D_1$ and $D_2$ with the quantum numbers
\beq
D_1:\,\,\, (1/2,0;1/2,1), \qquad D_2:\,\,\, (1/2,0;-1/2,-1)\,.
\label{dyons}
\eeq

We see that the quantum numbers of the massless quarks in the 
$r=2$ vacuum after the 
collision with the monopole singularity get shifted, the shift being equal to
$\pm$(monopole magnetic charge). 

The monodromy discussed above  implies 
\beq
a_1\to a_1 +a^3_D, \qquad a_2\to a_2 -a^3_D,\qquad a^3\to a^3 +2a^3_D\, .
\label{monodromy}
\eeq
Therefore, the conditions (\ref{masslessquarks}) for masslessness of the
 $q^{11}$ and $q^{22}$ quarks are  replaced in domain III by the conditions of 
masslessness of the dyons $D_1$ and $D_2$, namely,
\beq
a_1+ a^3_D+\frac{m_1}{\sqrt{2}}=0,\qquad a_2-a^3_D+\frac{m_2}{\sqrt{2}}=0\, .
\label{masslessdyons}
\eeq

\subsection{The low-energy theory}
\label{bulkIII}

In this subsection we present the low-energy theory 
in the $r=2$ vacuum in domain III at small $\xi$ and small $|\Delta m|$ 
(below the AD point). It should be stressed
that none of the fields in this low-energy theory belong to
nontrivial representations of SU(2)$_{C+F}$.

As we already know, the   massless quarks $q^{11}$and 
 $q^{22}$ transform into the massless dyons $D_1$ and $D_2$. The latter interact with
 two photons. According to the
dyon quantum numbers (\ref{dyons}) one of these photons is
\beq
A_{\mu},
\label{ph8}
\eeq
while the other photon is the following  linear combination:
\beq
B_{\mu}= \frac1{\sqrt{5}}\,(A^{3}_{\mu}+2A^{3D}_{\mu})\, .
\label{phB}
\eeq

In fact, these are the only light states to be included in the 
low-energy effective theory in domain III. All other states are either
heavy (with masses of the order of $\Lambda$) or decay on 
curves of marginal stability. In the case at hand CMS
 is located 
around the origin in the $\Delta m$ complex plane
and goes through the AD point \cite{svz}. In fact, the $W$ bosons
of the underlying non-Abelian gauge theory, as well as the off-diagonal
states of the quark matrix $q^{kA}$, decay on CMS. Let us illustrate this statement,
say, for the $W$ bosons. To this end we can go to the AD point.
At this point   we have  for the $W$-boson mass  
\beqn
m_W
&=&
\sqrt{2}|a^3|=\sqrt{2}\left|(a^3+a_D^3)-a^3_D\right|
\nonumber\\[2mm]
&=&
\sqrt{2}(\left|a^3+a_D^3\right|+\left| a^3_D\right|)=m_M +m_D\,,
\eeqn
where $m_M$ and $m_D$ are the masses of the SU(2) monopole and SU(2)
dyon with charges $(0,0;\, 0,1)$ and $(0,0;\,1,1)$, respectively. This relation is valid
at the AD point just because the monopole becomes massless at this
point, $a^3_D=0$. It means that the $W$-boson decays into the SU(2) monopole 
and dyon at this point and is not
present in domain III,  in full accordance with the analysis of
the SU(2) theory in \cite{BF}.

Taking this into account we write the effective low-energy
action of the theory in domain III as follows:
\beqn
S_{III}&=&\int d^4x \left[\frac{1}{4\tilde{g}^2_{2}}
\left(F^{B}_{\mu\nu}\right)^2 +
\frac1{4g^2_1}\left(F_{\mu\nu}\right)^2
+
\frac1{\tilde{g}^2_2}\left|\pt_{\mu}b\right|^2 +\frac1{g^2_1}
\left|\partial_{\mu}a\right|^2 \right.
\nonumber\\[4mm]
&+& \left|\nabla^1_{\mu}
D_1\right|^2 + \left|\nabla^1_{\mu} \tilde{D}_1\right|^2+
\left|\nabla^2_{\mu}
D_2\right|^2 + \left|\nabla^2_{\mu} \tilde{D}_{2}\right|^2
\nonumber\\[4mm]
&+&
\left. V(D,\tilde{D},b,a)\right]\,,
\label{SIII}
\eeqn
where 
\beq
b= \frac{1}{\sqrt{5}}\,(a^{3}+2a^{3}_D)
\label{fieldb}
\eeq
is the scalar \ntwo superpartner of the photon (\ref{phB})
while $F_{\mu\nu}^B$ is the field strength of the U(1) gauge field $B_{\mu}$. 
Covariant derivatives are defined in accordance
with the charges of the dyons $D_1$ and $D_2$. Namely,
\beqn
\nabla^1_\mu & = & \partial_\mu -i\left(\frac{1}{2}\; A_{\mu}
+ \frac12 A^{3}_{\mu}+A^{3D}_{\mu}\right) 
=\pt_{\mu}-\frac{i}{2}\left(A_{\mu}+\sqrt{5}B_{\mu}\right)\,,
\nonumber\\[3mm]
\nabla^2_\mu & = & \partial_\mu -i\left(\frac{1}{2}\; A_{\mu}
- \frac12 A^{3}_{\mu}-A^{3D}_{\mu}\right)
=\pt_{\mu}-\frac{i}{2}\left(A_{\mu}-\sqrt{5}B_{\mu}\right).
\label{nablaD}
\eeqn
The coupling constants $g_1$ and $\tilde{g}_2$
correspond to two U(1) gauge groups.
The potential $V(D,\tilde{D},b,a)$ in the action (\ref{SIII})
is 
\beqn
V(D,\tilde{D},b,a) &=&
 \frac{5\tilde{g}^2_2}{8}
\left( |D_1|^2 -
|\tilde{D}_1|^2 -|D_2|^2 +
|\tilde{D}_2|^2 \right)^2
\nonumber\\[3mm]
&+& \frac{g^2_1}{8}
\left(|D_1|^2 -|\tilde{D}_1|^2 +|D_2|^2 -
|\tilde{D}_2|^2 
-2 \xi\right)^2
\nonumber\\[3mm]
&+& \frac{5\tilde{g}_2^2}{2}\left| \tilde{D}_1 D_1-
\tilde{D}_2 D_2\right|^2+
\frac{g^2_1}{2}\left| \tilde{D}_1 D_1+
\tilde{D}_2 D_2 \right|^2
\nonumber\\[3mm]
&+&\frac12 \left\{ \left|a+\sqrt{5}b+\sqrt{2}m_1 
\right|^2\left(|D_1|^2+|\tilde{D}_1|^2\right)\right.
\nonumber\\[3mm]
&+&\left.
\left|a-\sqrt{5}b+\sqrt{2}m_2 
\right|^2\left(|D_2|^2+|\tilde{D}_2|^2\right) \right\}\,.
\label{potIII}
\eeqn

Now we are ready move to the desired limit of equal quark masses, $\Delta m=0$.
In this limit the global
SU(2)$_{C+F}$ symmetry is restored in the underlying theory. 
The vacuum of the 
theory  (\ref{SIII}) is
located at the following values of scalars $a$ and $b$:
\beq
a=-\sqrt{2}\,m,\qquad \sqrt{5}\,b=-\frac{\Delta m}{\sqrt{2}},
\label{abvev}
\eeq
while the VEVs of dyons are determined by the FI parameter $\xi$,
\beq
D_1=\sqrt{\xi},\qquad D_2=\sqrt{\xi},\qquad \tilde{D}_1=\tilde{D}_2=0\,.
\label{Dvev}
\eeq
Thus, the U(1)$\times$U(1) gauge group is broken by dyon condensation. Both, photons
and dyons become massive,  with masses proportional to $\sqrt{\xi}$.
In particular, at $\Delta m=0$  the vacuum value of $b$ vanishes. 

Note also that the theory (\ref{SIII}) is the Abelian U(1)$\times$U(1) gauge theory 
and hence is  not asymptotically free. It stays at weak coupling at small
$\xi$.

The low-energy theory (\ref{SIII}) does not seem to have any
global SU(2) symmetry. However, the underlying theory does have 
a global SU(2)
symmetry in the limit $\Delta m=0$. As was explained in Sect.~\ref{bulk},
this global SU(2) is not broken in domain I at large $\xi$. 
This symmetry is realized in a color-flavor-locked form  in 
this domain (see Eq.~(\ref{c+f})), and no Goldstone bosons are present. We showed 
 that no 
massless  states are present in domain III at non-zero $\xi$,
(and no light states other than two dyons and two photons discussed above); therefore,
the global SU(2) cannot be spontaneously broken in this domain either. The only
way out of this puzzle is to conclude
 that the SU(2) global symmetry is realized trivially in the low-energy description
(\ref{SIII}), i.e. that all states in (\ref{SIII}) are singlets of the unbroken
flavor SU(2).

This means, as was already mentioned in Sect.~\ref{smallerxi}, that the photon $B_{\mu}$
which appears in  domain III has nothing to do with the third component of the
SU(2) gauge field $A_{\mu}^a$ of domain I. At $\Delta m=0$ the former is a singlet
of the global SU(2), while the latter is a component of a triplet. 

Moreover, dyons $D_1$
and $D_2$ present in domain III have nothing to do with diagonal entries
of the quark matrix $q^{kA}$ of domain I. Dyons are singlets while the quarks $q^{kA}$
form the  singlet and triplet states. 

Since we have a crossover  between domains I and III rather than  a phase 
transition,
this means that in the full theory triplets become heavy and decouple as we pass from 
domain I into domain III along the line $\Delta m=0$. Moreover, some 
composite singlets, which are 
heavy  and invisible in domain I become light in   domain III and form dyons
$D_{1,2}$ and photon $B_{\mu}$ (level crossing). Although this crossover is smooth in the full theory,
from the standpoint of the low-energy description the passage from  domain I into 
domain III means a dramatic change: the low-energy theories in these domains are 
completely
different, in particular, the degrees of freedom in these theories are different
(non-Abelian in  domain I vs. Abelian in  domain III).

\subsection{Strings in domain III}
\label{sid3}

It is obvious that the low-energy theory (\ref{SIII}) have $Z_2$ string solutions
 in the vacuum
(\ref{abvev}), (\ref{Dvev}). Say, the $D_1$ dyon  can have a winding at infinity. In this
case the string solution has the following behavior at $r\to\infty$:
\beqn
D_1 & \sim  & e^{i\alpha}\sqrt{\xi},\qquad D_2\sim \sqrt{\xi},
\nonumber\\[3mm] 
A_i & \sim & \pt_i \alpha, \qquad \sqrt{5}\, B_i\sim \pt_i \alpha,
\label{strIII}
\eeqn
where the indices $i=1,2$ denote 
 the plane orthogonal to the string axis and $r$ and $\alpha$ are polar
coordinates in this plane.
Another elementary string can be obtained from the one in (\ref{strIII})
by the replacement $D_1\to D_2$, $D_2\to D_1$ and $B_i\to -B_i$.

These $Z_2$ elementary strings are BPS-saturated. Their
tensions   are  given by the  formula (\ref{ten}) in the same way as
the tensions of elementary strings in domains I and II.

The $Z_2$ strings are Abelian 
(of the Abrikosov--Nielsen--Olesen type \cite{ANO})  in domain III. They do not have any 
orientational moduli,
in contrast with non-Abelian strings in domain I. 

Let us calculate the gauge invariant 
non-Abelian flux (\ref{flux}) for these strings. In Abelian domain II at large 
$|\Delta m|$ 
\beq
a^a \,F^{*a}_{3}\to a^3 \,F^{*3}_{3}\, .
\eeq
With $|\Delta m|$ decreasing, as we pass through monodromies, we get
\beqn
\label{XXX1}
a^3&\to&
 a^3+2a^3_D=\sqrt{5}\,b\,,
\\[1mm]
A_{\mu}^3
&\to&
\sqrt{5}\,B_{\mu}\,.
\label{XXX2}
\eeqn
Equation~(\ref{XXX2})
follows from Eq.~(\ref{XXX1})  by \ntwo supersymmetry.
Therefore
\beq
\Phi_{\rm III}=\int d^2x (\sqrt{5}\,b)\,(\sqrt{5}\,F^{*B}_3)\,.
\eeq

Equation (\ref{abvev})  gives  $$\sqrt{5}\,b =-\Delta m/\sqrt{2}$$ in the
$r=2$ vacuum, while 
the flux of the field $B_{\mu}$ of the $Z_2$ strings can be read off from Eq.~(\ref{strIII}).
In this way we arrive at 
\beq
\langle\Phi  \rangle _{\rm III} \to \mp 2\pi\,\frac{\Delta m}{\sqrt{2}}\,.
\label{fluxIII}
\eeq

We see that the string flux in domain III is given by the same formula as in domain
 II. This is a flux of the Abelian string. The non-Abelian part of the flux
is directed in the Cartan subalgebra of the gauge group. No orientational moduli
appear.  In contrast, in domain I the flux of the non-Abelian string is
proportional to the orientational vector $S^a$. At small $|\Delta m|$ the expectation value
$\langle S^a\rangle \to 0$, and the string flux is averaged to zero, see (\ref{fluxI}).
Domains I and III are separated by a crossover at $\xi\sim\Lambda^2$.

Let us also mention one more dramatic distinction of nonperturbative spectra
in domains I and III at $\Delta m=0$. The confined SU(2) monopoles (with quantum numbers
(\ref{monopole}))
are the
junctions of two different elementary strings in both domains. In the non-Abelian
domain I the 
confined monopoles are seen as  kinks of the world sheet CP$(N-1)$ model   
 \cite{SYmon,T,HT2}. As was shown by Witten \cite{W79}, deep 
in the 
quantum regime at $(m_A-m_B)=0$ the kink of the CP$(N-1)$ model is 
described by the field
$n^l$ and therefore acquires a global flavor quantum number with respect to the 
unbroken SU$(N)_{C+F}$. In fact, the kink/monopole
 is in the fundamental representation of this group (a doublet in
the case $N=2$) \cite{W79,HoVa}. Therefore,  a meson formed by a monopole connected
to an antimonopole by two strings (see the review paper \cite{SYrev} for details)  belongs
to  the singlet or adjoint representations of the global SU$(N)$ (singlet or triplet of
 SU(2) for $N=2$).

Clearly, in domain III the monopole confined by strings does not acquire  global
quantum numbers. It is in the singlet representation of SU(2). Hence,
 a  meson formed by a monopole connected
to an antimonopole by two strings is a singlet too.
 Thus, in the nonperturbative spectra of the
 theory we observe the same phenomenon which was seen in the perturbative spectra:
triplets of global SU(2) present at low energies in domain I are lifted
and do not appear in the low-energy description in domain III. Both perturbative
and nonperturbative states in  domain III are singlets of the unbroken global  SU(2).

\section{The phase transition at \boldmath{$N\to\infty$}}
\label{largeN}
\setcounter{equation}{0}

In this section we will consider the $N$ dependence
of the  crossover transitions (see Fig.~\ref{figphasediag}) 
in parameters $\xi$ and $(m_A-m_B)$. We will show
that in the large-$N$ limit the crossovers become
exceedingly sharper and  at $N=\infty$ transform into genuine phase transition.
We will start from the crossover in $(m_A-m_B)$ at large $\xi$ (i.e. the passage from
 domain I to domain II).

This crossover in the nonperturbative sector of the theory
can be seen as a crossover in the effective CP$(N-1)$ model (\ref{mcpg})
on the world sheet of the non-Abelian string, see  Sect.~\ref{strings}.
As was already explained, at large quark mass differences the
CP$(N-1)$ model is at weak coupling. The VEV of the vector $n^l$ does not vanish.
If we make a special choice for the mass parameters
\beq
{m}_k= m_0 \,  e^{\frac{2\pi k}{N}i},\qquad k=1,...,N,
\label{masses}
\eeq
where $m_0$ is a single common parameter (which we will take to be real) our theory has a 
discrete $Z_{2N}$ symmetry, see Appendix for further details.
In fact, $\langle n^l\rangle$ is an order parameter for the spontaneous breaking of
 this  $Z_{2N}$ symmetry down to $Z_2$. 

At weak coupling,  at large $m_0$, dynamics  can be described as follows. The action
(\ref{mcpg}) contains a term
\beq
\left|\sigma-\frac{{m}_l}{\sqrt{2}}\right|^2\,|n^l|^2\,.
\label{sigman}
\eeq
At weak coupling the field $n$ can develop a VEV if $\sigma$ reduces to a particular
mass parameter,
\beq
\sigma=\frac{{m}_{k}}{\sqrt{2}},\qquad n^l=\sqrt{2\beta}\;\delta^{lk},
\label{vacuum}
\eeq
where $k=1,...,N$ labels $N$ different vacua (i.e. the elementary $Z_N$ strings of the bulk 
theory) and we rescaled the field $n^l$ in (\ref{mcpg}) to make its kinetic term canonic,
namely, $n^l\to n^l/\sqrt{2\beta}$.

As we reduce the value of $m_0$ the vacuum expectation value of 
the $n^l$ field becomes smaller and tends to zero at the left boundary of domain
II. Simultaneously, the VEV of $\sigma$ is no longer given by the mass, as in Eq.~(\ref{vacuum}).  In fact, $\sigma$
determines the bifermion condensate; $|\sigma|$ becomes of 
the order of $\Lambda_{\sigma}$ at 
$m_0\to 0$.
 In both limits the $Z_N$ symmetry is broken. This is the reason why two domains, I and II,
are separated by a crossover rather than a phase transition.\footnote{In the
nonsupersymmetric
case the VEV of $\sigma$ vanishes in the domain I, and the $Z_N$ symmetry is restored 
\cite{GSY05,GSYphtr}. In this case we do have a phase transition between domains I and II.}

In order to study the crossover at any $N$ (rather than at $N=\infty$)
we can use the description
of the supersymmetric CP$(N-1)$ model in terms of an exact superpotential 
\cite{W93,Dorey}. Upon integrating out $n^l$ fields the model can be described by an
exact twisted superpotential of the Veneziano--Yankielowicz type \cite{Veneziano} 
\beq
{\cal W}_{\rm eff}=\beta\;\Sigma + \frac1{4\pi}\sum_{l=1}^N\,
\left(\Sigma-\frac{{m}_l}{\sqrt{2}}\right)
\,\ln{\left(\Sigma-\frac{{m}_l}{\sqrt{2}}\right)},
\eeq
where $\Sigma$ is a twisted superfield \cite{W93} (with $\sigma$ being its lowest scalar
component) and we ignore here the $\theta$ dependence ($\theta$ stands for the vacuum angle). 
Minimizing this superpotential with 
respect to $\sigma$ we find
\beq
\prod_{l=1}^N(\sqrt{2}\,\sigma-{m}_l)=\Lambda_{\sigma}^N,
\label{sigmaeq}
\eeq
where $\Lambda_\sigma$ is the scale parameter of the CP$(N-1)$ sigma model under consideration.

Let us examine this  equation, determining the VEV of the field $\sigma$
at finite rather than infinite $N$. 
If $N$ is fixed, it is readily seen that at large $|{m}_l|$
 (i.e. $m_0\gg\Lambda_{\sigma}$) the solution for 
$\sigma$ coincides with one of the masses, in accordance with our semiclassical analysis,
see Eq.~(\ref{vacuum}). In the opposite limit of zero masses ($m_0=0$)
\beq
\sigma=\frac1{\sqrt{2}}\,\Lambda_{\sigma} e^{\frac{2\pi\,i\, k}{N}},
\label{sigmazeromass}
\eeq
where $ k=1,...,N$ marks $N$ distinct vacua. As was mentioned 
above, the  $Z_N$ symmetry is spontaneously broken at any ${m}_0$.

As we increase the value of $m_0$, the vacuum expectation of $\sigma$ smoothly interpolates between the regime 
(\ref{sigmazeromass}), where the order parameter which distinguishes different vacua
of the CP$(N-1)$ model (i.e. different elementary strings of the bulk theory) is a bifermion
condensate $\sim\sigma$, and the regime (\ref{vacuum}) where $\sigma$ is determined by one of the masses $ m_l$, while $n^l$ develops a VEV. For finite $N$ the solution for
$\sigma$ is a smooth function of $m_0$. Thus, this  is a crossover 
that takes place between  domains I and II. 

If we increase $N$ this crossover becomes more pronounced. Let us study 
Eq.~(\ref{sigmaeq}) at large $N$. To simplify our analysis let us consider
$N=2^p$, where $p$ is an integer. Then Eq. (\ref{sigmaeq}) can be rewritten as
\beq
(\sqrt{2}\sigma)^N-m_0^N=\Lambda_{\sigma}^N.
\eeq 
This equation has the following perfectly smooth solution:
\beq
\sigma=\frac1{\sqrt{2}}\,e^{\frac{2\pi k}{N}i}\,(m_0^N+\Lambda_{\sigma}^N)^{1/N}.
\label{sigmasol}
\eeq
However, at $N\to\infty$ the above function takes the form
\beq
\sigma=\frac1{\sqrt{2}}\,e^{\frac{2\pi k}{N}i}\,\times
\left\{
\begin{array}{l}\rule{0mm}{5mm}
m_0,\qquad  m_0>\Lambda_{\sigma}
\\[4mm]
 \Lambda_{\sigma},\qquad   m_0<\Lambda_{\sigma}
\end{array}
\right.\,.
\label{discontinuity}
\eeq
Corrections to this expression are
 exponential in $(-N)$.

We see that  the solution for $\sigma$ develops a discontinuity in the first derivative
with respect to $m_0$. The crossover becomes a phase transition in the limit $N=\infty$.
We stress that this phase transition is an artifact of the large-$N$ approximation
and is not related to a change in the pattern of realization  of any symmetry.

The solution (\ref{discontinuity}) for $\sigma$ ensures the following behavior of
the vector $n^l$ in the $N\to\infty$ limit:
\beq
\langle n^l\rangle=
\left\{
\begin{array}{l}\rule{0mm}{5mm}
\sqrt{2\beta_{\rm ren}}\;\delta^{kl},\qquad  m_0>\Lambda_{\sigma}\,,
\\[4mm]
 0,\qquad   m_0<\Lambda_{\sigma}\,,
\end{array}
\right.
\label{ndiscontinuity}
\eeq
where $k=1,...,N$.
The renormalized coupling $\beta_{\rm ren}$ tends to zero at $m_0=\Lambda_{\sigma}$ \cite{GSYphtr};
thus, the VEV of $n^l$ develops a discontinuity in the first derivative with respect to $m_0$.

This solution implies that the gauge invariant non-Abelian flux of the 
non-Abelian string
 strictly vanishes in domain I,
\beq
\langle \Phi\rangle_{\rm I}=0
\label{fluxIlargeN}
\eeq
at $N=\infty$ while in domains II and III it is given by an U$(N)$ generalization of 
Eq.~(\ref{fluxIII}). Namely, 
\beq
\langle \Phi\rangle_{\rm II}=\langle \Phi\rangle_{\rm III}
=-2\pi\,\sqrt{2}\,{m}_k
\eeq
for the $k$-th elementary $Z_N$ string, $k=1,...,N$, see Eqs.~(\ref{strflux})
and  (\ref{avev}).

The  result (\ref{fluxIlargeN}) is exact at $N=\infty$. Thus, at $N=\infty$ the string flux (\ref{flux})
develops a discontinuity as we pass from domain I to domains II or III.
This implies that
both crossovers in  $\xi$
and ${m}_0$ transform into phase transitions.

\section{Discussion}
\label{7seven}

In this paper we considered  the $r=N$ vacuum in \ntwo supersymmetric QCD with
the  U(N) gauge
group and $N_f=N$ flavors. We demonstrated that this theory exhibits a crossover
transition in $\xi$, see Fig~\ref{figphasediag}. Namely, at large $\xi$ 
in  domain I
the theory is in the non-Abelian confinement regime, it has 
$N^2$ degrees of freedom (gauge bosons and quarks) at low energies
and supports non-Abelian strings.  
In contrast, at small 
$\xi$ the theory passes into the Abelian Seiberg--Witten regime III. The low-energy effective 
description
includes $N$ degrees of freedom (dyons and dual photons) and supports Abelian strings.
We have shown that non-Abelian gauge bosons and quarks in domain I have nothing to
do with Abelian dyons and photons of  domain III. These states belong to different
representation of the unbroken global flavor group SU(N).

Although in this paper we considered a particular vacuum in a specially chosen
version of \ntwo SQCD (where the global SU$(N)$
symmetry remains unbroken due to the color-flavor locking)
we believe that our results are quite general. It  seems plausible that many
Abelian vacua of the Seiberg--Witten type in \ntwo SQCD exhibit  crossover transitions into 
non-Abelian regimes as we increase the FI parameter $\xi$. Usually we just do not have
appropriate extra parameters (such as the quark mass differences in our example) which 
would allow us to study these crossovers.

The lesson is that, generally speaking, non-Abelian strings
can smoothly evolve into Abelian strings
and vice versa.
At the same time the corresponding dynamical patterns are drastically different.

What conclusions apply to theories with less supersymmetry?
In the simplest version of the
Seiberg--Witten solution \cite{SW1},  \ntwo super\-symmetric QCD can be deformed by adding a mass term $\mu$ for the adjoint
field. In the limit of large $\mu$ the theory flows to \none SQCD. At small $\mu$ the mass term
for the adjoint field induces a Fayet--Iliopoulos $F$ term in \ntwo theory \cite{HSZ,VY},
 with
$\xi$ proportional to $\mu$ times some mass scale, such as $\Lambda$ or quark mass. Thus, the
deformation parameter $\mu$ translates, roughly speaking,  into the FI parameter $\xi$.

It is commonly believed that the behavior of supersymmetric QCD is smooth in
$\mu$: the Abelian degrees of freedom of \ntwo theory smoothly evolve into non-Abelian
degrees of freedom of \none theory as we increase $|\mu|$. While on the conceptual side
our results provide an unambiguous evidence in favor of the smooth transition,
dynamics-wise the emerging pictures on the opposite sides of domain lines separating
domains I, II and III hardly look alike.
 In particular, light degrees of freedom are completely different.
 
In addition we should note that there is at least one 
example of a nonsupersymmetric model where the evolution is proven to be
discontinuous,\footnote{On the other hand, analyses \cite{SU2,SU3}
carried out in a nonsupersymmetric setting 
different from that treated here and in \cite{GSY05,GSYphtr}
show no sign of the phase transition,
while \cite{SU4} exhibits a chiral phase transition on the way from 
Abelian to non-Abelian confinement.
} 
with a phase transition \cite{GSY05,GSYphtr}. And even in our basic \ntwo
model the 
crossover becomes a full-blown phase transition at $N=\infty$.

To conclude, we would like to comment on the recent paper \cite{Kabeliz}.
In this paper it is argued that non-Abelian vacua with $r>N_f/2$ which support non-Abelian 
strings
``dynamically Abelianize'' in quantum theory. We disagree with this statement.
 As we  demonstrated above,
both the Abelian and non-Abelian regimes can be present in \ntwo QCD in quantum theory. 
They just occur in 
different domains of the parameter space and are separated by   crossovers.

\section*{Acknowledgments}
This work  is supported in part by DOE grant DE-FG02-94ER408. 
The work of A.Y. was  supported 
by  FTPI, University of Minnesota, 
by RFBR Grant No. 09-02-00457a 
and by Russian State Grant for 
Scientific Schools RSGSS-11242003.2.

\vspace{2.5cm}

\section*{Appendix: \\
Global symmetries of the CP\boldmath{$(N-1)$} model with \boldmath{$Z_{2N}$}-symmetric
 twisted masses}
\addcontentsline{toc}{section}{Appendix: \\
Global symmetries of the CP$(N-1)$ model with  $Z_N$-symmetric
 twisted masses}
 \renewcommand{\theequation}{A.\arabic{equation}}
\setcounter{equation}{0}
 
 \renewcommand{\thesubsection}{A.\arabic{subsection}}
\setcounter{subsection}{0}

First, let us outline the \ntwo CP$(N-1)$ model with twisted masses
\cite{Alvarez} in one
of a few possible formulations, the so-called gauge formulation \cite{peregolo}.
This formulation is built on an $N$-plet of complex scalar fields $n^i$ where $i=1,2,...,N$.
We impose the constraint
\beq
n_i^\dagger\,n^i = 1\,.
\eeq
This leaves us with $2N-1$ real bosonic degrees of freedom. To eliminate one extra degree
of freedom we impose a local U(1) invariance $n^i(x)\to e^{i\alpha(x)} n^i(x)$.
To this end we introduce a gauge field $A_\mu$ which converts the partial derivative into the
covariant one,
\beq
\partial_\mu\to \nabla_\mu \equiv \partial_\mu -i\,  A_\mu\,.
\eeq
The field $A_\mu$ is auxiliary; it enters in the Lagrangian without derivatives. The kinetic term of the
$n$ fields is
\beq
L =\frac{2}{g_0^2}\, \left|\nabla_\mu n^i\right|^2\,.
\eeq

The superpartner to the field $n^i$ is an $N$-plet of complex two-component spinor fields $\xi^i$,
\beq
\xi^i =\left\{\begin{array}{l}
\xi^i_R\\[2mm]
\xi^i_L
\end{array}
\right.\,.
\eeq

The auxiliary field $A_\mu$ has a complex scalar superpartner $\sigma$ 
and a two-component complex spinor superpartner $\lambda$; both enter without derivatives.
The full \ntwo symmetric Lagrangian is
\beqn
L &=& \frac{2}{g_0^2}\left\{\rule{0mm}{6mm}
\left|\nabla_\mu n^i\right|^2+ \xi_i^\dagger\, i\gamma^\mu\nabla_\mu\,\xi^i
+ 2\sum_i\left|\sigma-\frac{m_i}{\sqrt 2}\right|^2\, |n^i|^2\right.
\nonumber\\[3mm]
&+&\left.\left[
i\sqrt{2}\,\sum_i \left( \sigma -\frac{m_i}{\sqrt 2}\right)\xi^\dagger_{iR}\, \xi^i_L +i\sqrt{2}\,n^\dagger_i
\left(\lambda_R\xi^i_L - \lambda_L\xi^i_R
\right)+{\rm H.c.}\right]
\right\}.
\nonumber\\
\label{bee31}
\eeqn
where $m_i$ are twisted mass parameters.
Equation (\ref{bee31}) is valid in a special case when
\beq
\sum_{i=1}^N \,m_i = 0\,.
\label{bee32}
\eeq
We will make a specific choice of the
parameters $m_i$, namely,
\beq
m_i= m  \left\{  e^{2\pi i/N} , \,e^{4\pi i/N},\, ... ,
 \,e^{2(N-1)\pi i/N} , \, 1\right\} \,,
\label{spch}
\eeq
where $m$ is a single common parameter. Then the constraint
(\ref{bee32}) is automatically satisfied. Without loss of generality $m$ can be assumed
to be real and positive. The U(1) gauge symmetry is built in. This symmetry eliminates one bosonic degree of freedom, leaving us with $2N-2$ dynamical bosonic degrees of freedom inherent to CP$(N-1)$ model.

Now let us discuss global symmetries of this model. In the absence of the twisted masses
the model was SU$(N)$ symmetric. The twisted masses (\ref{spch}) explicitly break this symmetry down to U$(1)^{N-1}$,
\beqn
n^\ell&\to& e^{i\alpha_\ell}n^\ell\,,\quad \xi^\ell_R \to e^{i\alpha_\ell}\xi^\ell_R\,
\quad \xi^\ell_L \to e^{i\alpha_\ell}\xi^\ell_L\,,\quad \ell=1,2, ..., N\,,
\nonumber\\[2mm]
\sigma
&\to&
 \sigma\,,\quad \lambda_{R,L}\to \lambda_{R,L}\,.
\eeqn
where $\alpha_\ell$ are $N$ constant phases different for different $\ell$. 

Next, there is a global vectorial U(1) symmetry which rotates all fermions $\xi^\ell$
in one and the same way, leaving the boson fields intact,
\beqn
\xi^\ell_R 
&\to& 
e^{i\beta}\xi^\ell_R\,, \quad
 \xi^\ell_L \to e^{i\beta}\xi^\ell_L\,,\quad \ell=1,2, ..., N\,,
\nonumber\\[2mm]
\lambda_R 
&\to&
 e^{-i\beta}\lambda_R\,,\quad 
\lambda_L \to e^{-i\beta}\lambda_L\,,
\nonumber\\[2mm]
n^\ell &\to& n^\ell\,,\quad \sigma\to\sigma\,.
\eeqn

Finally, there is a discrete $Z_{2N}$ symmetry which is of most importance for our purposes.
Indeed, let us start from the axial U$(1)_R$ transformation which would be a symmetry
of the classical action at $m=0$ 
 (it is anomalous, though, under quantum corrections),
\beqn
\xi^\ell_R 
&\to& 
e^{i\gamma}\xi^\ell_R\,, \quad
 \xi^\ell_L \to e^{-i\gamma }\xi^\ell_L\,,\quad \ell=1,2, ..., N\,,
 \nonumber\\[2mm]
 \lambda_R 
&\to&
 e^{i\gamma}\lambda_R\,,\quad 
 \lambda_L \to e^{-i\gamma}\lambda_L\,,\quad \sigma \to e^{2i\gamma}\sigma\,,
\nonumber\\[2mm]
n^\ell
&\to&
 n^\ell\,.
\eeqn
With $m$ switched on and the chiral anomaly included, this transformation 
is no longer the symmetry of the model. However, a discrete $Z_{2N}$ subgroup survives both the inclusion of anomaly and $m\neq 0$. This subgroup corresponds to
\beq
\gamma_k =\frac{2\pi i k}{2N}\,,\quad k= 1,2, ..., N\,.
\eeq
with the simultaneous shift
\beq
\ell\to \ell - k\,.
\eeq
In other words,
\beqn
\xi^\ell_R 
&\to& 
e^{i\gamma_k}\xi^{\ell-k}_R\,, \quad
 \xi^\ell_L \to e^{-i\gamma_k }\xi^{\ell-k}_L\,, 
 \nonumber\\[2mm]
 \lambda_R 
&\to&
 e^{i\gamma_k}\lambda_R\,,\quad 
 \lambda_L \to e^{-i\gamma_k}\lambda_L\,,\quad \sigma \to e^{2i\gamma_k}\sigma\,,
 \nonumber\\[2mm]
 n^\ell &\to & n^{\ell-k}\,.
 \label{bee35}
\eeqn
This $Z_{2N}$ symmetry  relies on the particular choice of masses 
given in (\ref{spch}).

The order parameters for the $Z_N$ symmetry are as follows:
(i) the set of the vacuum expectation values
$\{ \langle n^1\rangle,\,\, \langle n^2\rangle, \,...\, \langle n^N\rangle\}$
and (i) the bifermion condensate $\langle  \xi^\dagger_{R,\,\ell}\xi^\ell_L\rangle$.
Say, a nonvanishing value of $\langle n^1\rangle$ or  $\langle  \xi^\dagger_{R,\,\ell}\xi^\ell_L\rangle$ implies that the $Z_{2N}$ symmetry of the action is broken down to
$Z_2$. The first order parameter is more convenient for detection
at large $m$ while the second at small $m$. 

It is instructive to illustrate the above conclusions
in a different formulation of the sigma model, namely, in the geometrical formulation
(for simplicity we will consider CP(1); generalization to CP$(N-1)$ is straightforward).
In components the Lagrangian of the model is
\beq
\begin{split}
{\cal L}_{\,  CP(1)}&= G\, \Big\{
\partial_\mu \phi^\dagger\, \partial^\mu\phi -|m|^2{\phi^\dagger\,\phi} 
+\frac{i}{2}\big(\psi_L^\dagger\!\stackrel{\leftrightarrow}{\partial_R}\!\psi_L 
+ \psi_R^\dagger\!\stackrel{\leftrightarrow}{\partial_L}\!\psi_R
\big)
\\[1mm] 
&
-i\,\frac{1-\phi^\dagger\,\phi}{\chi} \,\big(m\,\psi_L^\dagger \psi_R + \bar m
\psi_R^\dagger \psi_L
\big)
\\[1mm] 
&
-\frac{i}{\chi}\,  \big[\psi_L^\dagger \psi_L
\big(\phi^\dagger \!\stackrel{\leftrightarrow}{\partial_R}\!\phi
\big)+ \psi_R^\dagger\, \psi_R
\big(\phi^\dagger\!\stackrel{\leftrightarrow}{\partial_L}\!\phi
\big)
\big]
\\[1mm]
&
-
\frac{2}{\chi^2}\,\psi_L^\dagger\,\psi_L \,\psi_R^\dagger\,\psi_R
\Big\}\,,
\end{split}
\label{Aone}
\eeq
where 
\beq
\chi = 1+\phi^\dagger\,\phi\,,\quad G= \frac{2}{g_0^2\,\chi^2}
\eeq
and 
\beq
\partial_L =\frac{\partial}{\partial t} +\frac{\partial}{\partial z}\,,\qquad
\partial_R =\frac{\partial}{\partial t} - \frac{\partial}{\partial z}\,.
\eeq
The $Z_2$ transformation corresponding to (\ref{bee35}) is
\beq
\phi \to -\frac{1}{\phi^\dagger}\,,\qquad \psi_R^\dagger \psi_L\to -
\psi_R^\dagger \psi_L\,.
\label{bee40}
\eeq
 The order parameter which can detect breaking/nonbreaking of the above
symmetry is
\beq
\frac{m}{g_0^2} \left(1- \frac{g_0^2}{2\pi}
\right)\, \frac{\phi^\dagger\,\phi-1}{\phi^\dagger\,\phi+1} - 
i R \psi_R^\dagger \psi_L\,.
\eeq
Under the transformation (\ref{bee40}) this order parameter changes sign.
In fact, this is  the central charge of the \ntwo
sigma model, including the anomaly  \cite{svz}.

Now, what changes if instead of the \ntwo model we will consider nonsupersymmetric 
CP$(N-1)$ model with twisted masses? Then the part of the Lagrangian (\ref{bee31}) containing fermions must be dropped. The same must be done in the $Z_2$ order parameter.
As was shown in~\cite{GSY05,GSYphtr}, now at $m>\Lambda$ the $Z_2$ symmetry is broken, while at $m<\Lambda$ unbroken. A phase transition takes place.

\vspace{2.5cm}



\small
\begin{thebibliography}{99}
\itemsep -2pt

\bibitem{SW1}
N.~Seiberg and E.~Witten,
Nucl. Phys. {\bf B426}, 19 (1994),
(E) {\bf B430},  485 (1994) [hep-th/9407087].

\bibitem{SU2}
M.~Shifman and M.~\"Unsal,
  Phys.\ Rev.\  D {\bf 78}, 065004 (2008)
  [arXiv:0802.1232 [hep-th]].
  
  \bibitem{SU3}
  M.~Shifman and M.~\"Unsal,
 {\em On Yang--Mills Theories with Chiral Matter at Strong Coupling,}
  arXiv:0808.2485 [hep-th].

  \bibitem{Pol}
A.~M.~Polyakov,
  Nucl.\ Phys.\  B {\bf 120}, 429 (1977).
  
  \bibitem{GSY05}
A.~Gorsky, M.~Shifman and A.~Yung,
  Phys.\ Rev.\ D {\bf 71}, 045010 (2005)
  [hep-th/0412082].
  
  \bibitem{GSYphtr}
A. Gorsky, M. Shifman, and A. Yung,
Phys. \ Rev.\  D {\bf 73} (2006) 065011
[hep-th/0512153].

\bibitem{SYrev}
M.~Shifman and A.~Yung,
{\sl Supersymmetric Solitons,}
Rev.\ Mod.\ Phys. {\bf 79} 1139 (2007)
[arXiv:hep-th/0703267].
  
\bibitem{HT1}
A.~Hanany and D.~Tong,
JHEP {\bf 0307}, 037 (2003)
[hep-th/0306150].

\bibitem{ABEKY}
R.~Auzzi, S.~Bolognesi, J.~Evslin, K.~Konishi and A.~Yung,
Nucl.\ Phys.\ B {\bf 673}, 187 (2003)
[hep-th/0307287].

 \bibitem{SYmon}
M.~Shifman and A.~Yung,
Phys.\ Rev.\ D {\bf 70}, 045004 (2004)
[hep-th/0403149].

\bibitem{HT2}
A.~Hanany and D.~Tong,
JHEP {\bf 0404}, 066 (2004)
[hep-th/0403158].

\bibitem{Trev}
D.~Tong,
{\em TASI Lectures on Solitons,}
  arXiv:hep-th/0509216.

\bibitem{Jrev}
  M.~Eto, Y.~Isozumi, M.~Nitta, K.~Ohashi and N.~Sakai,
  J.\ Phys.\ A  {\bf 39}, R315 (2006)
  [arXiv:hep-th/0602170].

\bibitem{Trev2}
D.~Tong
{\em Quantum Vortex Strings: A Review,}
arXiv:0809.5060 [hep-th].
  
\bibitem{SW2}
N.~Seiberg and E.~Witten,
Nucl. Phys. {\bf B431}, 484  (1994)
[hep-th/9408099].

\bibitem{VY}
A.~I.~Vainshtein and A.~Yung,
Nucl.\ Phys.\ B {\bf 614}, 3 (2001)
[hep-th/0012250].

\bibitem{W93}
E.~Witten,
  Nucl.\ Phys.\ B {\bf 403}, 159 (1993)
  [hep-th/9301042].
  
    \bibitem{Po3}
A.~M.~Polyakov,
Phys.\ Lett.\ B {\bf 59}, 79 (1975).

\bibitem{W79}
E.~Witten,
Nucl.\ Phys.\ B {\bf 149}, 285 (1979).

  \bibitem{NSVZsigma}
V.~A.~Novikov, M.~A.~Shifman, A.~I.~Vainshtein and
V.~I.~Zakharov,
Phys.\ Rept.\  {\bf 116}, 103 (1984).

\bibitem{Alvarez}
L.~Alvarez-Gaum\'e and D.~Z.~Freedman,
Commun.\ Math.\ Phys.\  {\bf 91}, 87 (1983);
S.~J.~Gates,
Nucl.\ Phys.\ B {\bf 238}, 349 (1984);
S.~J.~Gates, C.~M.~Hull and M.~Ro\v{c}ek,
Nucl.\ Phys.\ B {\bf 248}, 157 (1984).

\bibitem{Dorey}
N.~Dorey,
JHEP {\bf 9811}, 005 (1998) [hep-th/9806056].

\bibitem{ArFa}
P. C.~Argyres and A. E.~Faraggi,
Phys. \ Rev. \ Lett. {\bf 74},  3931 (1995)
[hep-th/9411057].

\bibitem{KLTY}
A.~Klemm, W.~Lerche, S.~Yankielowicz and S.~Theisen,
Phys. \ Lett. \ B {\bf 344}, 169 (1995) 
[hep-th/9411048].

\bibitem{ArPlSh}
P. C. Argyres, M. R. Plesser, and  A. Shapere,
Phys. \ Rev. \ Lett.  \ {\bf 75}, 1699 (1995)
[hep-th/9505100].

\bibitem{HaOz}
A.~Hanany and  Y.~Oz,
Nucl. \ Phys. \ B {\bf 452}, 283 (1995)
[hep-th/9505075].

\bibitem{BF}
A.~Bilal and F.~Ferrari,
  Nucl.\ Phys.\  B {\bf 516}, 175 (1998)
  [arXiv:hep-th/9706145].

\bibitem{CKM}
G.~Carlino, K.~Konishi and H.~Murayama,
Nucl.\ Phys.\ B {\bf 590}, 37 (2000)
[hep-th/0005076].
 
\bibitem{AD}
P. C.~Argyres and M. R.~Douglas,
Nucl. \ Phys. \ {\bf B448}, 93 (1995)   
[arXiv:hep-th/9505062].
  
\bibitem{APSW}
P. C. Argyres, M. R. Plesser, N. Seiberg, and E. Witten,
Nucl. \ Phys.  \ {\bf B461}, 71 (1996) 
[arXiv:hep-th/9511154].

  \bibitem{svz}
  A.~Losev and M.~Shifman,
Phys.\ Rev.\ D {\bf 68}, 045006 (2003)
[hep-th/0304003];
M.~Shifman, A.~Vainshtein and R.~Zwicky,
  J.\ Phys.\ A  {\bf 39}, 13005 (2006)
  [arXiv:hep-th/0602004].

\bibitem{ANO}
A.~Abrikosov, Sov.~Phys. JETP {\bf32}, 1442  (1957)
[Reprinted in {\em Solitons and Particles}, Eds. C. Rebbi and G. Soliani
(World Scientific, Singapore, 1984), p. 356];\\
H.~Nielsen and P.~Olesen, Nucl.~Phys. {\bf B61}, 45 (1973)
[Reprinted in {\em Solitons and Particles}, Eds. C. Rebbi and G. Soliani
(World Scientific, Singapore, 1984), p. 365].

  \bibitem{T}
D.~Tong,
Phys.\ Rev.\ D {\bf 69}, 065003 (2004)
[hep-th/0307302].

\bibitem{HoVa}
K.~Hori and C.~Vafa,
{\em Mirror Symmetry,}
hep-th/0002222.

\bibitem{Veneziano}
  G.~Veneziano and S.~Yankielowicz,
  Phys.\ Lett.\  B {\bf 113}, 231 (1982).

\bibitem{HSZ}
A.~Hanany, M.~J.~Strassler and A.~Zaffaroni,
Nucl.\ Phys.\ B {\bf 513}, 87 (1998)
[hep-th/9707244].

\bibitem{SU4}
M. Shifman, M. \"Unsal,
{\em Multiflavor QCD$^*$ on $R_3 \times S_1$: Studying Transition From Abelian to Non-Abelian Confinement},
arXiv:0901.3743 [hep-th].

\bibitem{Kabeliz}
D. Dorigoni, K. Konishi, and K. Ohashi,
{\em Non-Abelian Vortices without Dynamical Abelianization,}
arXiv:0801.3284 [hep-th].

\bibitem{peregolo}
H.~Eichenherr,
  Nucl.\ Phys.\  B {\bf 146}, 215 (1978)
  [Erratum-ibid.\  B {\bf 155}, 544 (1979)];
  V.~L.~Golo and A.~M.~Perelomov,
  Lett.\ Math.\ Phys.\  {\bf 2}, 477 (1978);
  E.~Cremmer and J.~Scherk,
  Phys.\ Lett.\  B {\bf 74}, 341 (1978).
 


\end{thebibliography}
\end{document}